\documentclass[12pt]{article}

\def\comp{{\rm C}\llap{\vrule height7.1pt width1pt depth-.4pt\phantom t}}
\def\Fint{\rlap{$\Biggl\rfloor$}\Biggl\lceil}
\def\square{\kern1pt\vbox{\hrule height 1.2pt\hbox{\vrule width 1.2pt\hskip 3pt
   \vbox{\vskip 6pt}\hskip 3pt\vrule width 0.6pt}\hrule height 0.6pt}\kern1pt}

\begin{document}

\begin{titlepage}

\begin{flushright}
SPIN-09/19, ITP-UU-09/19 \\ CRETE-09-15 \\ UFIFT-QG-09-05
\end{flushright}

\vskip 1cm

\begin{center}
{\bf Transforming to Lorentz Gauge on de Sitter}
\end{center}

\vskip .5cm

\begin{center}
S. P. Miao$^*$
\end{center}

\begin{center}
\it{Institute for Theoretical Physics \& Spinoza Institute, Utrecht
University \\ Leuvenlaan 4, Postbus 80.195, 3508 TD Utrecht, THE NETHERLANDS}
\end{center}

\vskip .3cm

\begin{center}
N. C. Tsamis$^{\dagger}$
\end{center}

\begin{center}
\it{Department of Physics, University of Crete \\
GR-710 03 Heraklion, HELLAS}
\end{center}

\vskip .3cm

\begin{center}
R. P. Woodard$^{\ddagger}$
\end{center}

\begin{center}
\it{Department of Physics, University of Florida \\
Gainesville, FL 32611, UNITED STATES}
\end{center}

\vspace{.3cm}

\begin{center}
ABSTRACT
\end{center}
We demonstrate that certain gauge fixing functionals cannot be added to
the action on backgrounds such as de Sitter in which a linearization 
instability is present. We also construct the field dependent gauge
transformation which carries the electromagnetic vector potential from
a convenient, non-de Sitter invariant gauge to the de Sitter invariant, 
Lorentz gauge. The transformed propagator agrees with the de Sitter 
invariant result previously found by solving the propagator equation in
Lorentz gauge. This shows that the gauge transformation technique will 
eliminate unphysical breaking of de Sitter invariance introduced by a 
gauge condition. It is suggested that the same technique can be used to 
finally resolve the issue of whether or not free gravitons are de Sitter 
invariant.

\begin{flushleft}
\end{flushleft}

\begin{flushleft}
$^*$ e-mail: S.Miao@uu.nl \\
$^{\dagger}$ e-mail: tsamis@physics.uoc.gr \\
$^{\ddagger}$ e-mail: woodard@phys.ufl.edu
\end{flushleft}

\end{titlepage}

\section{Introduction}

Working out propagators is the difficult part about formulating
quantum field theoretic perturbation theory on exotic backgrounds.
It is typically accomplished by solving the differential equation
that the propagator must obey, however, this procedure is ambiguous
up to a homogeneous solution. It has long been realized that some
choices for this homogeneous solution do not make the resulting
Green's function into a true propagator. That is, the Green's
function does not correspond to the expectation value of the
time-ordered product of two free fields in the presence of any state
\cite{TW1}.

There is nothing mysterious about the problem, nor does it require
field theory to understand. Consider the simple harmonic oscillator
whose position as a function of time is $q(t)$ and whose Lagrangian
is,
\begin{equation}
L = \frac12 m \dot{q}^2 - \frac12 m \omega^2 q^2 \; .
\end{equation}
The propagator equation for this system is,
\begin{equation}
-m \Bigl[ \Bigl(\frac{d}{dt}\Bigr)^2 + \omega^2\Bigr] i\Delta(t;t')
= i \delta(t \!-\! t') \; . \label{HOprop}
\end{equation}
The general solution to this equation which is symmetric under
interchange of $t$ and $t'$ has three free parameters,
\begin{eqnarray}
\lefteqn{i\Delta(t;t') = -\frac{i}{2 m \omega} \, \sin\Bigl[\omega
\vert t \!-\! t'\vert\Bigr] + \alpha \cos(\omega t) \cos(\omega t')}
\nonumber \\
& & \hspace{5.5cm} + \beta \sin\Bigl[\omega (t \!+\! t')\Bigr] +
\gamma \sin(\omega t) \sin(\omega t') \; . \qquad \label{gensol}
\end{eqnarray}
Although any choice of $\alpha$, $\beta$ and $\gamma$ in
(\ref{gensol}) gives a solution to the propagator equation
(\ref{HOprop}), the result is not a propagator unless they obey two
inequalities,
\begin{equation}
\alpha + \gamma \geq \frac1{2 m \omega} \qquad {\rm and} \qquad
\alpha \gamma \geq \frac14 \beta^2 \; . \label{ineqs}
\end{equation}
To see this, note first that the Heisenberg picture operator $q(t)$
can be expressed in terms of its initial position and momentum as,
\begin{equation}
q(t) = q_0 \cos(\omega t) + \frac{p_0}{m \omega} \, \sin(\omega t)
\; .
\end{equation}
For $i\Delta(t;t')$ to be a propagator there must be a state $\vert
\psi\rangle$ such that,
\begin{eqnarray}
\lefteqn{i\Delta(t;t') = \Bigl\langle \psi \Bigl\vert T\Bigl[ q(t)
q(t') \Bigr] \Bigr\vert \psi \Bigr\rangle } \\
& & = -\frac{i}{2 m \omega} \, \sin\Bigl[\omega \vert t \!-\!
t'\vert\Bigr] + \Bigl\langle \psi \Bigl\vert \frac{q_0^2}{2}
\Bigr\vert \psi \Bigl\rangle \,
\cos(\omega t) \cos(\omega t') \nonumber \\
& & \hspace{.3cm} + \Bigl\langle \psi \Bigl\vert \frac{q_0 p_0 \!+\!
p_0 q_0}{2 m \omega} \Bigr\vert \psi \Bigr\rangle \,
\sin\Bigl[\omega (t \!+\! t')\Bigr] + \Bigr\langle \psi \Bigl\vert
\frac{p_0^2}{2 m^2 \omega^2} \Bigr\vert \psi \Bigr\rangle \,
\sin(\omega t) \sin(\omega t') \; . \qquad
\end{eqnarray}
We can therefore identify the constants $\alpha$, $\beta$ and
$\gamma$ as,
\begin{equation}
\alpha = \Bigl\langle \psi \Bigl\vert \frac{q_0^2}{2} \Bigr\vert
\psi \Bigl\rangle \quad , \quad \beta = \Bigl\langle \psi \Bigl\vert
\frac{q_0 p_0 \!+\! p_0 q_0}{2 m \omega} \Bigr\vert \psi
\Bigl\rangle \quad , \quad \gamma = \Bigl\langle \psi \Bigl\vert
\frac{p_0^2}{2 m^2 \omega^2} \Bigr\vert \psi \Bigl\rangle \; .
\end{equation}
For the ground state one has,
\begin{equation}
{\rm Ground\ State} \qquad \Longrightarrow \qquad \alpha = \gamma =
\frac1{4 m \omega} \quad {\rm and} \quad \beta = 0 \; .
\end{equation}
For a general state the first inequality in (\ref{ineqs}) results
from requiring the expectation value of the energy to be greater
than or equal to $\frac12 \omega$; the second is just the Schwarz
inequality.

A more subtle set of issues can arise in gauge theories. To
understand the one of interest for this work we must digress to
explain the difference between an ``exact gauge'' and an ``average
gauge'' \cite{TW2}. The former is obtained by choosing the gauge
parameter to make the vector potential obey some equation at each
point in space and time. This is the normal type of gauge fixing in
classical field theory. Familiar examples are,
\begin{eqnarray}
{\rm Temporal\ Gauge} & : & A_0(t,\vec{x}) = 0 \; , \\
{\rm Coulomb\ Gauge} & : & \vec{\nabla} \!\cdot\! \vec{A}(t,\vec{x})
= 0
\; , \\
{\rm Lorentz\ Gauge} & : & \partial^{\mu} A_{\mu}(t,\vec{x}) =
-\dot{A}_0(t,\vec{x}) + \vec{\nabla} \!\cdot\! \vec{A}(t,\vec{x}) =
0 \; .
\end{eqnarray}
Although exact gauges can be used in quantum field theory the more
common type of gauge fixing is accomplished by adding some
noninvariant term to the invariant Lagrangian. For example, the
Feynman gauge Lagrangian is,
\begin{equation}
\mathcal{L} = -\frac14 F_{\mu\nu} F^{\mu\nu} - \frac12
(\partial^{\mu} A_{\mu})^2 \; . \label{Feyn}
\end{equation}
The functional integral representation for this type of gauge
condition can be viewed as a weighted average of exact gauges. For
example, the Feynman gauge functional formalism results from
imposing the exact gauge,
\begin{equation}
\partial^{\mu} A_{\mu}(t,\vec{x}) = f(t,\vec{x}) \; ,
\end{equation}
where $f(x)$ is a $\comp$-number field. One then functionally
averages over $f(x)$ with a Gaussian weighting functional,
\begin{equation}
\Fint [df] \, \exp\Bigl[-\frac{i}2 \int \!\! dt \int \!\! d^3x \,
f^2(t,\vec{x}) \Bigr] \; .
\end{equation}

From this discussion it is obvious that a fairly involved set of
functional changes of variables connects the exact gauge conditions
of the canonical formalism to the average gauge conditions typically
employed in the functional formalism. The late Sidney Coleman worked
this out explicitly for flat space on the manifold $R^4$ to derive
the Faddeev-Popov ansatz for this case \cite{SRC}, but the result is
often assumed without justification for general metrics on any
manifold. We suspect that the unjustified use of average gauge
fixing is behind a dispute about the graviton propagator on de
Sitter background.

It is easy to see that certain average gauges can be problematic
when linearization instabilities are present. Consider flat space
electrodynamics on the manifold $T^3 \times R$. Because the spatial
sections are compact, both sides of the spatially averaged, $\mu = 0$ 
Maxwell equation must vanish separately,
\begin{equation}
\partial_{\nu} F^{\nu\mu} = J^{\mu} \qquad \Longrightarrow \qquad
\int \!\! d^3x \, \partial_i F^{i0}(t,\vec{x}) = \int \!\! d^3x \,
J^0(t,\vec{x}) \; .
\end{equation}
Because this zero charge constraint follows from the invariant field
equations it must be true as well in every exact gauge. However,
naively imposing an average gauge can result in a very different
theory. For example the field equations of Feynman gauge
(\ref{Feyn}) are,
\begin{equation}
\Bigl[-\partial_t^2 + \nabla^2\Bigr] A^{\mu}(t,\vec{x}) =
J^{\mu}(t,\vec{x}) \; .
\end{equation}
These equations can be solved for any total charge. One can argue
about how the problem happened, or how significant it is, but there
cannot be any doubt that something went wrong.

The issues we have been discussing are relevant to a debate between
cosmologists and relativists concerning perturbative quantum gravity
on de Sitter background. From the perspective of inflationary
cosmology it is natural to view de Sitter as a special case of
homogeneous, isotropic and spatially flat geometries whose invariant
element in co-moving coordinates takes the general form,
\begin{equation}
ds^2 = -dt^2 + a^2(t) d\vec{x} \cdot d\vec{x} \; .
\end{equation}
One gets the open coordinate sub-manifold of de Sitter by setting
the scale factor to $a(t) = e^{Ht}$ with constant $H$. For any scale
factor the transverse-traceless components of the graviton field
obey the same equation as the massless, minimally coupled scalar
\cite{Grishchuk},
\begin{equation}
\Bigl[\Bigl(\frac{\partial}{\partial t}\Bigr)^2 + 3 H
\frac{\partial}{\partial t} - \frac{\nabla^2}{a^2}\Bigr]
h^{tt}_{ij}(t,\vec{x}) = 0 \; .
\end{equation}
The power spectrum for gravitational radiation \cite{AAS1} is
proportional to the canonically normalized, super-horizon scalar
mode functions $u(t,k) \sim H/k^{\frac32}$,
\begin{equation}
\mathcal{P}_h \sim G \times \vert u(t,k)\vert^2 \times k^3 \sim G
H^2 \; .
\end{equation}
From scale invariance --- which would be exact in de Sitter
--- one sees that the mode functions of physical gravitons diverge
too strongly at small $k$ to give a convergent result for the
Fourier mode sum of a part of the graviton propagator which must be
present in any gauge. It follows that there can be no de Sitter
invariant graviton propagator, just as there is no de Sitter
invariant propagator for the massless, minimally coupled scalar
\cite{AF}.

People who abhor the breaking of de Sitter invariance typically
dismiss it as unphysical, but this argument cannot be accepted
because the tensor power spectrum is certainly physical. (Indeed,
strenuous efforts are underway to observe it!) Nor is there any
support to be gained from the tiny distinction between de Sitter and
primordial inflation, which typically makes the infrared behavior
worse in any case. So one would think that the noninvariance of free
gravitons on de Sitter must be accepted as universally as that of
the massless, minimally coupled scalar. This is is not so because
relativists have been able to find de Sitter invariant solutions to
the propagator equation in average gauges \cite{math}. Explicit
solutions in what seem to be valid gauges are just as compelling as
inferences from the tensor power spectrum. However, we have just
seen that average gauges may not be reliable on manifolds such as de
Sitter which possess linearization instabilities.

The early recognition of a problem \cite{IAEM} with one of these de
Sitter invariant solutions led to the development of a noninvariant,
average gauge \cite{TW3,RPW1}. The associated propagator has been
shown to obey the Ward identities at tree order \cite{TW4} and one
loop \cite{TW5}; and the only fully renormalized, dimensionally
regulated loop results for quantum gravity on de Sitter background
have been obtained using it \cite{TW6,MW,KW1}. Although the gauge
fixing functional of this propagator is not de Sitter invariant, it
does preserve the one-parameter subgroup of dilatations,
\begin{eqnarray}
t & \longrightarrow & t - \frac1{H} \, \ln(k) \; , \\
\vec{x} & \longrightarrow & k \vec{x} \; ,
\end{eqnarray}
so the fact that the propagator breaks dilatation invariance cannot
be blamed on the gauge. Moreover, Kleppe has shown that the
propagator's de Sitter breaking is physical by the standard
technique of supplementing naive de Sitter transformations with a
compensating gauge transformation to restore the noninvariant gauge
condition \cite{Kleppe}.

As with the cosmological power spectrum, one would think these
results decisive, but the interest in a de Sitter invariant graviton
propagator persists \cite{Higuchi}. What seems to be necessary to
settle the issue is two things:
\begin{itemize}
\item{A proof that the average gauges for which de Sitter invariant
solutions have been found are not valid; and}
\item{An explicit construction of the graviton propagator in an exact,
de Sitter invariant gauge which is valid over the full de Sitter
manifold.}
\end{itemize}
Of course the imposition of a de Sitter invariant gauge would make
the propagator equation de Sitter invariant, but that does not imply
a de Sitter invariant solution for the graviton propagator any more
than it does for the massless, minimally coupled scalar propagator
which obeys an invariant equation but is not invariant \cite{AF}. If
the graviton propagator in an exact gauge breaks de Sitter
invariance then the breaking must be accepted as physical. This
would not only resolve a contentious dispute, the resulting
propagator might be easier to use and it would reduce the number of
counterterms \cite{MW,KW1}.

In section 2 of this paper we give a proof that the average, de Sitter 
invariant gauges are not valid; in subsequent sections we develop the 
machinery for constructing the propagator in exact, de Donder gauge. 
The technique for our construction is to find the field-dependent gauge 
transformation $\xi_{\mu}[h]$ which enforces the
gauge condition,
\begin{equation}
h_{\mu\nu}' = h_{\mu\nu} - 2 \xi_{\mu ; \nu}[h] \qquad {\rm such\
that} \qquad g^{\rho\sigma} \Bigl[h_{\mu \rho ; \sigma}' - \frac12
h_{\rho\sigma ; \mu}'\Bigr] = 0 \; .
\end{equation}
(In this and subsequent formulae, $g_{\mu\nu}$ stands for the
spacelike background de Sitter metric and a semi-colon denotes
covariant differentiation with respect to this background.) Then we
use this transformation to carry the existing, noninvariant
propagator \cite{TW3,RPW1} into de Donder gauge. If the de Sitter
breaking of the existing propagator is completely due to the
noninvariant gauge then the de Donder gauge result should be
invariant; if the de Sitter breaking is physical then it should
persist after the transformation.

As a warmup for de Sitter gravity we here carry out the same exercise 
for de Sitter electromagnetism. That is, we find the field-dependent
gauge transformation $\theta[A]$ which imposes exact Lorentz gauge,
\begin{equation}
A_{\mu}' = A_{\mu} - \partial_{\mu} \theta[A] \qquad {\rm such\
that} \qquad \partial_{\mu} \Bigl( \sqrt{-g} g^{\mu\nu}
A_{\nu}'\Bigr) = 0 \; .
\end{equation}
Then we use this transformation on the photon propagator in a
noninvariant, average gauge \cite{RPW1,KW2}. Because we already know
the unique, de Sitter invariant solution for the propagator equation
in Lorentz gauge \cite{TW7}, obtaining this known solution by
transformation demonstrates that the technique will remove de Sitter
breaking that arises from using a noninvariant gauge condition. The
simpler setting of electromagnetism, and the close relation between
the noninvariant electromagnetic and gravitational gauges should
also teach us much of value for the main project.

This paper consists of six sections of which the first is ending.
In section 2 we review the functional changes of variables which
carry one from an exact, canonical gauge to a covariant, average
gauge, with special attention to what goes wrong when linearization
instabilities are present. The context is flat space electrodynamics
on the $D$-dimensional manifolds $R^D$ and $T^{D-1} \!\times\! R$. In
section 3 we switch to $D$-dimensional de Sitter and carry out the
field-dependent gauge transformation that enforces exact, Lorentz
gauge. Of course this gauge transformation is ambiguous up to a 
homogeneous solution, which is uniquely determined by de Sitter
invariance but which we leave unspecified at this stage. In section 
4 we describe the unique, de Sitter invariant solution that was 
found by solving the propagator equation in Lorentz gauge \cite{TW7}.
In section 5 we show how the hitherto unspecified, homogeneous part 
of the gauge transformation from section 3 can be chosen to make the 
two propagators agree. Our conclusions comprise section 6.

\section{Deriving Average Gauges}

The purpose of this section is to explain how one derives average
gauges from the exact gauges of the canonical formalism. We shall
take flat space quantum electrodynamics (QED) as the object of study
and first sketch the technique for passing from Coulomb gauge to
Feynman gauge on the manifold $R^D$. We then consider the same
process for the manifold $T^{D-1} \times R$ to show explicitly why
full Feynman gauge cannot be imposed.

The QED Lagrangian is,
\begin{equation}
\mathcal{L} = -\frac14 F_{\mu\nu} F^{\mu\nu} + \overline{\psi}
\gamma^{\mu} \Bigl(i \partial_{\mu} \!-\! e A_{\mu}\Bigr) \psi - m
\overline{\psi} \psi \; .
\end{equation}
The canonical dynamical variables of Coulomb gauge ($\vec{\nabla}
\cdot \vec{A}(t,\vec{x}) = 0$) are the transverse components of the
vector potential, $A_i^T(t,\vec{x})$, and the electric field,
$E_i^T(t,\vec{x})$, as well as $\psi(t,\vec{x})$ and
$\overline{\psi}(t,\vec{x})$. The $\mu = 0$ component of the vector
potential is not an independent dynamical variable but rather a
nonlocal functional $\Phi[\overline{\psi},\psi](t,\vec{x})$ of the
charged fields,
\begin{equation}
A_0 = \Phi[\overline{\psi},\psi] \equiv -\frac1{\nabla^2} \Bigl[e
\overline{\psi} \gamma^0 \psi\Bigr] \; .
\end{equation}
And the Hamiltonian density of Coulomb gauge is,
\begin{equation}
\mathcal{H} = \frac12 E_i^T E_i^T + \frac14 F_{ij} F_{ij} - \frac12
\Phi \nabla^2 \Phi + \overline{\psi}\Bigl( -i \gamma^i \partial_i +
e \gamma^i A_i^T + m \Bigr) \psi \; .
\end{equation}

The usual connection between the canonical and functional formalisms
is made through the vacuum expectation values of $T^*$-ordered
products of operators. ($T^*$-ordering is the same as time-ordering
except that derivatives are taken outside the ordering.) Suppose
$O[\overline{\psi},\psi,E^T,A^T]$ represents an arbitrary functional
of the canonical operators. The fundamental relation between the
canonical and functional integral formalisms is,
\begin{eqnarray}
\lefteqn{\Bigl\langle \Omega \Bigl\vert T^*\Bigl(
\mathcal{O}\Bigl[\overline{\psi},\psi,E^T,A^T\Bigr]\Bigr) \Bigr\vert
\Omega \Bigr\rangle } \nonumber \\
& & \hspace{1cm} = \Fint [d\overline{\psi}] [d\psi] [dE^T] [dA^T] \,
e^{i \int \! d^Dx \, [i\overline{\psi} \gamma^0 \dot{\psi} + E^T_i
\dot{A}^T_i - \mathcal{H}]} \, \mathcal{O}\Bigl[ \overline{\psi},
\psi,E^T,A^T\Bigr] \; . \qquad \label{FI1}
\end{eqnarray}
Because they will play no role in our analysis we have suppressed
the initial and final state wave functionals.

We henceforth denote expression (\ref{FI1}) as $\langle
\mathcal{O}\rangle$. A 7-step process of functional manipulations
carries it to Feynman gauge on $R^D$:
\begin{enumerate}
\item{{\it Perform the Gaussian integrals over the transverse
electric field,}
\begin{eqnarray}
\lefteqn{\Fint [dE^T] \, e^{i \int d^Dx \, [E^T_i \dot{A}^T_i -
\frac12 E^T_i E^T_i]} \,
\mathcal{O}\Bigl[\overline{\psi},\psi,E^T,A^T\Bigr]} \nonumber \\
& & \hspace{5.6cm} =
\mathcal{O}'\Bigl[\overline{\psi},\psi,A^T\Bigr] e^{i\int \! d^Dx \,
\frac12 \dot{A}^T_i \dot{A}^T_i} \; , \qquad \label{FI2}
\end{eqnarray}
where $\mathcal{O}'[\overline{\psi},\psi,A^T]$ is
$\mathcal{O}[\overline{\psi},\psi,\dot{A}^T,A^T]$ plus the delta
function correlator terms that arise from eliminating the various
pairings of $E^T$'s.}
\item{{\it Restore the longitudinal component of the vector potential},
\begin{equation}
\Fint [dA^T] = \Fint [d\vec{A}] \, \delta\Bigl[\vec{\nabla}
\!\cdot\! \vec{A}\Bigr] \, \sqrt{{\rm det}(-\nabla^2)} \;.
\label{FI3}
\end{equation}
Note that this gives the square root of the Faddeev-Popov
determinant for Coulomb gauge.}
\item{{\it Restore the temporal component of the vector potential,}
\begin{equation}
e^{i \int \! d^Dx \, \frac12 \Phi \nabla^2 \Phi} = \sqrt{{\rm
det}(-\nabla^2)} \times \Fint [dA_0] \, e^{i \int \! d^Dx \,
[\frac12 \partial_i A_0 \partial_i A_o - \overline{\psi} \gamma^0 e
A_0 \psi]} \; . \label{FI4}
\end{equation}
Note that this gives the remaining bit of the Faddeev-Popov
determinant. At this stage $\langle \mathcal{O}\rangle$ takes the
form,
\begin{eqnarray}
\lefteqn{\langle \mathcal{O}\rangle = \Fint [d\overline{\psi}]
[d\psi] [dA] \, \delta\Bigl[ \vec{\nabla}\! \cdot \! \vec{A}\Bigr]
\, {\rm det}(-\nabla^2) } \nonumber \\
& & \hspace{-.3cm} \times e^{i\int \! d^Dx \, [\frac12 \dot{A}_i
\dot{A}_i + \frac12 \partial_i A_0 \partial_i A_0 - \frac14 F_{ij}
F_{ij} + \overline{\psi} (i \gamma^{\mu} \partial_{\mu} -
\gamma^{\mu} e A_{\mu} - m ) \psi]} \, \mathcal{O''}\Bigl[
\overline{\psi},\psi,A\Bigr] \; , \qquad \label{FI5}
\end{eqnarray}
where $\mathcal{O}''[\overline{\psi},\psi,A]$ is
$\mathcal{O}'[\overline{\psi},\psi,\vec{A}]$ with possible factors
of $\Phi[\overline{\psi},\psi]$ replaced by $A_0$(if desired; it is
not necessary) and with the addition of appropriate correlator terms
for pairings of $A_0$'s.}
\item{{\it Express the integrand as an invariant.} Any good gauge
can be used to express an arbitrary functional of the fields as a
gauge invariant which happens to agree with the original functional
when the gauge condition is obeyed \cite{TW2,RPW2}. We do this for
the action and for the operator
$\mathcal{O}''[\overline{\psi},\psi,A]$,
\begin{eqnarray}
\delta\Bigl[\vec{\nabla} \!\cdot\! \vec{A}\Bigr] \times e^{i \int \!
d^Dx \, [\frac12 \dot{A}_i \dot{A}_i + \frac12 \partial_i A_0
\partial_i A_0]} = \delta\Bigl[\vec{\nabla} \! \cdot \!
\vec{A}\Bigr] \times e^{i \int \! d^Dx \, \frac12 F_{0i} F_{0i}} \;
, \label{FI6a} \\
\delta\Bigl[\vec{\nabla} \!\cdot\! \vec{A}\Bigr] \times
\mathcal{O}''\Bigl[\overline{\psi},\psi,A\Bigr] =
\delta\Bigl[\vec{\nabla} \!\cdot\! \vec{A}\Bigr] \times
\mathcal{O}_{\rm inv}\Bigl[\overline{\psi},\psi,A\Bigr] \; .
\label{FI6b}
\end{eqnarray}
After invariantizing in this way our expression for $\langle
\mathcal{O}\rangle$ is,
\begin{equation}
\langle \mathcal{O}\rangle = \Fint [d\overline{\psi}] [d\psi] [dA]
\, \delta\Bigl[\vec{\nabla} \!\cdot\! \vec{A}\Bigr] \, {\rm
det}(-\nabla^2) \, e^{i S_{\rm inv}} \, \mathcal{O}_{\rm inv} \; .
\label{FI7}
\end{equation}}
\item{{\it Make a functional change of variables to Lorentz gauge.}
Consider the field-dependent gauge transformation,
\begin{eqnarray}
A_{\mu}' & = & A_{\mu} - \partial_{\mu} \theta_1[A] \; , \\
\psi' & = & e^{ie \theta_1[A]} \times \psi \; , \\
\overline{\psi}' & = & e^{-ie \theta_1[A]} \times \overline{\psi} \;
\end{eqnarray}
where the gauge parameter is,
\begin{equation}
\theta_1[A] = -\frac1{\partial^2} \, \dot{A}_0 \; .
\end{equation}
Because $S_{\rm inv}$ and $\mathcal{O}_{\rm inv}$ are gauge
invariant, only the gauge fixing delta functional and the measure
will change. To get them, note that the inverse transformation for
the vector potential is,
\begin{equation}
A_{\mu} = A_{\mu} ' -\partial_{\mu} \frac1{\nabla^2} \dot{A}_0' \; .
\end{equation}
It follows that the Coulomb gauge condition on $A_{\mu}$ implies the
Lorentz gauge condition on $A_{\mu}'$,
\begin{equation}
\partial_i A_i = \partial^{\mu} A_{\mu} ' \; . \label{AtoA'}
\end{equation}
And the functional Jacobian converts the Faddeev-Popov determinant
to the one appropriate for Lorentz gauge,
\begin{equation}
[dA] \, {\rm det}(-\nabla^2) = [dA'] \, {\rm det}(-\partial^2) \; .
\end{equation}
We can therefore write $\langle \mathcal{O} \rangle$ as,
\begin{equation}
\langle \mathcal{O}\rangle = \Fint [d\overline{\psi}'] [d\psi']
[dA'] \, \delta\Bigl[\partial^{\mu} A_{\mu}'\Bigr] \, {\rm
det}(-\partial^2) \, e^{i S_{\rm inv}} \, \mathcal{O}_{\rm inv} \; .
\label{FI8}
\end{equation}}
\item{{\it Add a inhomogeneous, $\comp$-number term to the gauge
fixing functional.} Make an additional change of variable,
\begin{eqnarray}
A_{\mu}'' & = & A_{\mu}' - \partial_{\mu} \theta_2[f] \; , \\
\psi'' & = & e^{i e \theta_2[f]} \times \psi \; , \\
\overline{\psi}'' & = & e^{-i e \theta_2[f]} \times \overline{\psi}
\; ,
\end{eqnarray}
where the gauge parameter is defined in terms of an arbitrary
$\comp$-number field $f(t,\vec{x})$,
\begin{equation}
\theta_2[f] = \frac1{\partial^2} \, f \; .
\end{equation}
After the transformation we have,
\begin{equation}
\langle \mathcal{O}\rangle = \Fint [d\overline{\psi}''] [d\psi'']
[dA''] \, \delta\Bigl[\partial^{\mu} A_{\mu}'' - f\Bigr] \, {\rm
det}(-\partial^2) \, e^{i S_{\rm inv}} \, \mathcal{O}_{\rm inv} \; .
\label{FI9}
\end{equation}}
\item{{\it Functionally average over the inhomogeneous term.}
By construction $\langle \mathcal{O} \rangle$ has no dependence upon
the function $f(t,\vec{x})$. It is therefore unchanged if we
multiply by a normalized Gaussian and functional integrate over $f$,
\begin{eqnarray}
\langle \mathcal{O} \rangle & = & \Fint [df] \, e^{i \int \! d^Dx \,
\frac12 f^2} \times \langle \mathcal{O}\rangle \; , \\
& = & \Fint [d\overline{\psi}''] [d\psi''] [dA''] \, {\rm
det}(-\partial^2) \, e^{i S_{\rm Feynman}} \, \mathcal{O}_{\rm inv}
\; . \qquad \label{FI10}
\end{eqnarray}
This is the Feynman gauge functional formalism we sought.}
\end{enumerate}

Let us see how the sequence of functional manipulations described above
changes when the noncompact spatial manifold $R^{D-1}$ is replaced by the 
compact manifold $T^{D-1}$. A major difference is that Fourier integrals 
become discrete sums. Suppose the range of each spatial coordinate is $-L 
\leq x^i < +L$ for $i = 1,2,\ldots,(D\!-\!1)$. Then any function 
$f(t,\vec{x})$ can be expressed as a discrete Fourier sum,
\begin{equation}
f(t,\vec{x}) = \sum_{\vec{n} \in Z^{D-1}} f_{\vec{n}}(t) e^{i k
\vec{n} \cdot \vec{x}} \; ,
\end{equation}
where $k \equiv \pi/L$ is the fundamental wave number. Note that the action 
of $\nabla^2$ on any such function annihilates the $\vec{n} = 0$ mode,
\begin{equation}
\nabla^2 f(t,\vec{x}) = - \sum_{\vec{n} \in Z^{D-1}} (k n)^2
f_{\vec{n}}(t) e^{i\pi L^{-1} \vec{n} \cdot \vec{x}} \; . 
\end{equation}
Hence the instantaneous Coulomb potential can only be defined for 
configurations of $\overline{\psi}(t,\vec{x})$ and $\psi(t,\vec{x})$ which
have zero total charge (this is the linearization stability constraint!) and
the resulting potential has no $\vec{n} = 0$ mode,
\begin{eqnarray}
\Phi(t,\vec{x}) & = & -\frac1{\nabla^2} \Bigl[\overline{\psi} \gamma^0 e 
\psi\Bigr](t,\vec{x}) \; , \\
& = & \sum_{\vec{n} \neq 0} \frac1{(k n)^2 (2L)^{D-1}} \int \!\! d^{D-1}\!x' 
\, e^{i k \vec{n} \cdot (\vec{x} - \vec{x}')} \; \overline{\psi}(t,\vec{x}') 
\gamma^0 e \psi(t,\vec{x}') \; . \qquad
\end{eqnarray}
Of course this means that when $A_0(t,\vec{x})$ is restored in step 3, it 
cannot contain any $\vec{n} = 0$ mode.
 
Another important change concerns restoring the longitudinal part of 
the vector potential in step 2. Under a gauge transformation the Fourier
mode $\vec{A}_{\vec{n}}(t)$ goes to,
\begin{equation}
\vec{A}_{\vec{n}}'(t) = \vec{A}_{\vec{n}}(t) - i k n \theta_{\vec{n}}(t) \; .
\end{equation}
It follows that all three vector components of $\vec{A}_0(t)$ are physical.
(In the noninteracting theory they would behave like free quantum mechanical 
particles rather than harmonic oscillators.) A second consequence is that 
the Coulomb gauge delta functional lacks a $\vec{n} = 0$ mode,
\begin{equation}
\delta\Bigl[\vec{\nabla} \!\cdot\! \vec{A}\Bigr] = \prod_{t \in R} 
\prod_{\vec{n} \neq 0} \delta\Bigl( k \vec{n} \!\cdot\! \vec{A}_{\vec{n}}(t)
\Bigr) \; .
\end{equation}

To recapitulate, the changes associated with working on the compact spatial
manifold $T^{D-1}$ are:
\begin{itemize}
\item{The field $A_0(t,\vec{x})$ contains no $\vec{n} = 0$ mode; and}
\item{The Coulomb gauge delta functional contains no $\vec{n} = 0$ mode.}
\end{itemize}
It is immediately obvious that, while we can still make the functional 
change of variables in step 7 to enforce exact Lorentz gauge, neither the
resulting $A_0'(t,\vec{x})$ nor the Lorentz gauge delta functional will 
contain an $\vec{n} = 0$ mode. {\it It follows that there is no valid way
to add the $\vec{n} = 0$ mode of the Feynman gauge fixing term.} So the
result is just what we expected: adding the full Feynman gauge fixing
term is incorrect, and all conclusions drawn from this formalism are suspect.
It should be clear that the same sort of problem must occur as well in de
Sitter, and for gravitons as well as for electromagnetism, {\it so we now 
have a proof that the average gauges for which de Sitter invariant solutions 
have been found are not valid.} 

Let us return to the context of flat space electromagnetism on the manifold
$T^{D-1} \!\times \! R$ and consider what goes wrong if the problem we have 
just demonstrated is ignored and the covariant gauge fixing term (with 
$\vec{n} = 0$ mode) is erroneously added to the invariant action. In that 
case there is an extra, homogeneous contribution to $A_0(t,\vec{x})$, which 
should not be present, and a corresponding extra contribution to propagator. 
For a point charge $q$ this extra term produces a homogeneous contribution 
to $A_0(t,\vec{x})$ which grows like $t^2$,
\begin{equation}
A_0(t,\vec{x}) = \frac{q t^2}{2^D L^{D-1}} + {\rm Higher\ Modes} \; .
\end{equation}
One might object that the extra term is harmless because it makes no
contribution to the electric field, however, the undifferentiated potential
does contribute to the interaction energy. A special case of some interest
in quantum electrodynamics is the interaction of a particle with its own force
fields. In this context it has been noted that using the de Sitter invariant, 
Feynman gauge propagator \cite{AJ} (which must also contain a spurious 
homogeneous part) results in on-shell singularities for the one loop 
self-mass-squared of a charged scalar \cite{KW2}. Just as the analysis of 
this section suggests, these on-shell singularities disappear either:
\begin{itemize}
\item{When using a non-de Sitter invariant gauge on the open coordinate 
sub-manifold (which has no linearization instability) \cite{KW2}; or}
\item{When using the de Sitter invariant Lorentz propagator (which is exact) 
\cite{PTW1}.}
\end{itemize}
It should be emphasized that there is no mistake in the Allen-Jacobson 
solution for the Feynman gauge propagator \cite{AJ}; it is the gauge fixing
functional which is at fault.

\section{Imposing Lorentz Gauge on de Sitter}

The purpose of this section is to make a field-dependent gauge transformation
which carries the photon propagator from a non-de Sitter invariant average
gauge, defined on the open coordinate submanifold, to the de Sitter invariant,
exact Lorentz gauge, which can be extended to the entire de Sitter manifold.
We begin by reviewing the geometry and coordinate system, then we give the
noninvariant gauge condition and the associated propagator. The next step is
making the transformation. Of course this is ambiguous up to surface terms
which we leave to be specified in section 5. We close by decomposing the 
transformed propagator (without the homogeneous contributions) up into two 
convenient pieces.

We work on the open conformal coordinate submanifold of $D$-dimensional de 
Sitter space. A spacetime point $x^{\mu}$ can be decomposed into its temporal
($x^0$) and spatial $x^i$ components which take values in the ranges,
\begin{equation}
-\infty < x^0 < 0 \qquad {\rm and} \qquad {\rm and} -\infty < x^i < 
+\infty \; .
\end{equation}
In these coordinates the invariant element is,
\begin{equation}
ds^2 \equiv g_{\mu\nu} dx^{\mu} dx^{\nu} = a_x^2 \eta_{\mu\nu} dx^{\mu} 
dx^{\nu} \; ,
\end{equation}
where $\eta_{\mu\nu}$ is the Lorentz metric and $a_x = -1/Hx^0$ is the
scale factor. The parameter $H$ is known as the ``Hubble constant''.

Most of the various propagators between points $x^{\mu}$ and
$z^{\mu}$ can be expressed in terms of the de Sitter length function
$y(x;z)$,
\begin{equation}
y(x;z) \equiv \Bigl\Vert \vec{x} \!-\! \vec{z}\Bigr\Vert^2 - \Bigl(
\vert x^0 \!-\! z^0\vert \!-\! i \epsilon\Bigr)^2 \; . \label{ydef}
\end{equation}
Except for the factor of $i\epsilon$ (whose purpose is to enforce Feynman 
boundary conditions) the function $y(x;z)$ is closely related to the
invariant length $\ell(x;z)$ from $x^{\mu}$ to $z^{\mu}$,
\begin{equation}
y(x;z) = 4 \sin^2\Bigl( \frac12 H \ell(x;z)\Bigr) \; .
\end{equation}
Because $y(x;z)$ is a de Sitter invariant, so too are covariant derivatives
of it,
\begin{eqnarray}
\frac{\partial y(x;z)}{\partial x^{\mu}} & = & H a_x \Bigl(y \delta^0_{\mu} 
\!+\! 2 a_z H \Delta x_{\mu} \Bigr) \; , \\
\frac{\partial y(x;z)}{\partial z^{\nu}} & = & H a_z \Bigl(y \delta^0_{\nu} 
\!-\! 2 a_x H \Delta x_{\nu} \Bigr) \; , \\
\frac{\partial^2 y(x;z)}{\partial x^{\mu} \partial z^{\nu}} & = & H^2 a_x a_z 
\Bigl(y \delta^0_{\mu} \delta^0_{\nu} \!+\! 2 a_z H \Delta x_{\mu} 
\delta^0_{\nu} \!-\! 2 a_x \delta^0_{\mu} H \Delta x_{\nu} \!-\! 2 
\eta_{\mu\nu}\Bigr) \; . \qquad
\end{eqnarray}
Here and subsequently $\Delta x_{\mu} \equiv \eta_{\mu\nu} (x \!-\! z)^{\nu}$.

Electromagnetism is conformally invariant in $D=4$ dimensions, which means
that it takes the same form in conformal coordinates as in flat space. This 
is obvious from the gauge invariant Lagrangian,
\begin{equation}
\mathcal{L}_{\rm inv} = -\frac14 F_{\mu\nu} F_{\rho\sigma} g^{\mu\rho}
g^{\sigma\nu} \sqrt{-g} = -\frac14 a^{D-4} F_{\mu\nu} F_{\rho\sigma} 
\eta^{\mu\rho} \eta^{\nu\sigma} \; .
\end{equation}
The wonderful simplicity of using known flat space results will not be 
preserved if one adds any multiple of the de Sitter invariant, Feynman
gauge fixing functional,
\begin{equation}
\mathcal{L}_{\rm dS} = -\frac12 \Bigl(g^{\mu\nu} A_{\mu ; \nu}\Bigr)^2
= -\frac12 a^{D-4} \Bigl(\eta^{\mu\nu} A_{\mu , \nu} - (D\!-\!2) H a A_0
\Bigr)^2 \; . \label{LdS}
\end{equation}
(A semi-colon denotes covariant differentiation whereas a comma stands
for the ordinary derivative.) However, a very simple formalism results 
from replacing the factor of $(D\!-\!2)$ with $(D\!-\!4)$,
\begin{equation}
\mathcal{L}_{\rm NdS} = -\frac12 a^{D-4} \Bigl(\eta^{\mu\nu} \partial_{\mu} 
A_{\nu} - (D \!-\! 4) H a A_0\Bigr)^2 \; . \label{LNdS}
\end{equation}
With this gauge fixing functional the propagator takes the form 
\cite{RPW1,KW2},
\begin{equation}
i\Bigl[\mbox{}_{\mu} \Delta^{\rm NdS}_{\nu}\Bigr](x;z) = a_x a_z 
i\Delta_B(x;z) \Bigl(\eta_{\mu\nu} + \delta^0_{\mu} \delta^0_{\nu}\Bigr)
- a_x a_z i\Delta_C(x;z) \delta^0_{\mu} \delta^0_{\nu} \; , \label{NdSprop}
\end{equation}
where the de Sitter invariant scalar propagators are,
\begin{eqnarray}
i\Delta_B(x;z) & \equiv & B\Bigl(y(x;z)\Bigr) = 
\frac{H^{D-2}}{(4\pi)^{\frac{D}2}} \frac{\Gamma(D \!-\!2)}{\Gamma(\frac{D}2)}
\, \mbox{}_2 F_1\Bigl(D\!-\!2,1;\frac{D}2;1 \!-\! y\Bigr) \; , \qquad \\
i\Delta_C(x;z) & \equiv & C\Bigl(y(x;z)\Bigr) = 
\frac{H^{D-2}}{(4\pi)^{\frac{D}2}} \frac{\Gamma(D \!-\!3)}{\Gamma(\frac{D}2)}
\, \mbox{}_2 F_1\Bigl(D\!-\!3,2;\frac{D}2;1 \!-\! y\Bigr) \; . \qquad
\end{eqnarray}

One nice thing about (\ref{NdSprop}) is that its tensor factors are constants.
Another is that each of the scalar propagators which multiply them consists
of the conformal propagator plus a series of less singular terms which vanish
in $D=4$ dimensions,
\begin{eqnarray}
\lefteqn{B(y) = \frac{H^{D-2}}{(4\pi)^{\frac{D}2}} \Biggl\{\Gamma\Bigl(
\frac{D}2 \!-\! 1\Bigr) \Bigl(\frac4{y}\Bigr)^{\frac{D}2 -1} + 
\sum_{n=0}^{\infty} \Biggr[ \frac{\Gamma(n \!+\! \frac{D}2)}{\Gamma(n \!+\! 2)}
\Bigl(\frac{y}4 \Bigr)^{n -\frac{D}2 + 2} } \nonumber \\
& & \hspace{6.7cm} - \frac{\Gamma(n\!+\! D\!-\!2)}{\Gamma(n\!+\! \frac{D}2)} 
\Bigl(\frac{y}4\Bigr)^{n}\Biggr]\Biggr\} \; , \qquad \label{Bexp} \\
\lefteqn{C(y) = \frac{H^{D-2}}{(4\pi)^{\frac{D}2}} \Biggl\{\Gamma\Bigl(
\frac{D}2 \!-\! 1\Bigr) \Bigl(\frac4{y}\Bigr)^{\frac{D}2 -1} - 
\sum_{n=0}^{\infty} \Biggr[ \Bigl(n \!-\! \frac{D}2 \!+\!3 \Bigr) 
\frac{\Gamma(n \!+\! \frac{D}2 \!-\! 1)}{\Gamma(n \!+\! 2)} 
\Bigl(\frac{y}4 \Bigr)^{n-\frac{D}2+2} } \nonumber \\
& & \hspace{6.7cm} - (n \!+\! 1) \frac{\Gamma(n\!+\! 
D\!-\!3)}{\Gamma(n\!+\!  \frac{D}2)} \Bigl(\frac{y}4\Bigr)^{n}\Biggr]\Biggr\} 
\; . \qquad \label{Cexp}
\end{eqnarray}
So the infinite series terms only need to be retained when they multiply
a potentially divergent quantity. Because the higher values of $n$ vanish
more and more rapidly at coincidence (that is, for $y = 0$) only a finite 
number of these extra terms ever needs to be included.

It is now time to make the field-dependent transformation to Lorentz gauge,
\begin{equation}
A_{\mu}'(x) = A_{\mu}(x) - \partial_{\mu} \theta[A](x) \; .
\end{equation}
This would be step 5 in the scheme of the previous section. The fact that
$A_{\mu}'$ obeys Lorentz gauge implies a differential equation for $\theta[A]$,
\begin{equation}
\partial_{\mu} \Bigl( \sqrt{-g} g^{\mu\nu} \partial_{\nu} \theta\Bigr) =
\partial_{\mu} \Bigl( \sqrt{-g} g^{\mu\nu} A_{\nu}\Bigr) \; .
\end{equation}
Of course there are many solutions, related to one another by homogeneous 
terms. Any choice of homogeneous term will enforce Lorentz gauge, whereas
there can be at most one choice which gives a de Sitter invariant propagator.
Because Section 5 is devoted to establishing de Sitter invariance and
correspondence with the known solution \cite{TW7}, we postpone specification
of the homogeneous term until then. For now we express the solution in a
general way,
\begin{equation}
\overline{\theta}[A](x) = \int_{V} \! d^Dx' \, G(x;x') \frac{\partial}{
\partial x^{\prime \rho}} \Bigl( \sqrt{-g(x')} g^{\rho\sigma}(x') 
A_{\sigma}(x') \Bigr) \; , \label{thetabar}
\end{equation}
where $G(x;x')$ is some Green's function of the scalar d'Alembertian, which 
we specify in the next section, and $V$ is some region of the manifold. The 
actual solution for $\theta[A](x)$ consists of (\ref{thetabar}) --- with
definite choices for $G(x;x')$ and $V$ --- plus a definite homogeneous
solution. For now we study the field transformed with only 
$\overline{\theta}[A](x)$,
\begin{equation}
\overline{A}_{\mu}(x) \equiv A_{\mu}(x) - \partial_{\mu} 
\overline{\theta}[A](x) \; .
\end{equation}

The transformed propagator is the vacuum expectation value of the
$T^*$-ordered product of two $\overline{A}$'s. Because $T^*$-ordering
moves any derivatives outside the time-ordering symbol we can express this
as,
\begin{eqnarray}
\lefteqn{\Bigl\langle \Omega \Bigl\vert T^*\Bigl[ \overline{A}_{\mu}(x)
\overline{A}_{\nu}(z)\Bigr] \Bigr\vert \Omega \Bigr\rangle =
\Bigl\langle \Omega \Bigl\vert T\Bigl[ A_{\mu}(x) A_{\nu}(z)\Bigr] \Bigr\vert 
\Omega \Bigr\rangle } \nonumber \\
& & \hspace{-.5cm} -\frac{\partial}{\partial x^{\mu}} \int_{V} \! d^Dx' \,
G(x;x') \times \frac{\partial}{\partial x^{\prime \rho}} \Biggl[ \sqrt{-g(x')}
g^{\rho\sigma}(x') \Bigl\langle \Omega \Bigl\vert T\Bigl[ A_{\sigma}(x') 
A_{\nu}(z)\Bigr] \Bigr\vert \Omega \Bigr\rangle \Biggr] \nonumber \\
& & \hspace{-.5cm} -\frac{\partial}{\partial z^{\nu}} \int_{V} \! d^Dz' \,
G(z;z') \times \frac{\partial}{\partial z^{\prime \alpha}} \Biggl[ \sqrt{-g(z')}
g^{\alpha\beta}(z') \Bigl\langle \Omega \Bigl\vert T\Bigl[ A_{\mu}(x) 
A_{\beta}(z')\Bigr] \Bigr\vert \Omega \Bigr\rangle \Biggr] \nonumber \\
& & \hspace{-.5cm} + \frac{\partial}{\partial x^{\mu}} \frac{\partial}{
\partial z^{\nu}} \int_{V} \! d^Dx' \, G(x;x') \int_{V} \! d^Dz' \, G(z;z')
\nonumber \\
& & \times \frac{\partial}{\partial x^{\prime \rho}} 
\frac{\partial}{\partial z^{\prime \alpha}} \Biggl[ \! \sqrt{-g(x')} 
g^{\rho\sigma}(x') \sqrt{-g(z')} g^{\alpha\beta}(z')
\Bigl\langle \Omega \Bigl\vert T\Bigl[ A_{\sigma}(x') 
A_{\beta}(z')\Bigr] \Bigr\vert \Omega \Bigr\rangle \! \Biggr] . \qquad
\end{eqnarray}
By substituting the noninvariant propagator (\ref{NdSprop}) and using the
fact that $y(x;z)$ depends upon spatial coordinates only through their
difference, we can write the three differentiated, square-bracketed terms as,
\begin{eqnarray}
\lefteqn{\frac{\partial}{\partial x^{\prime \rho}} \Biggl[ \sqrt{-g(x')}
g^{\rho\sigma}(x') \Bigl\langle \Omega \Bigl\vert T\Bigl[ A_{\sigma}(x') 
A_{\nu}(z)\Bigr] \Bigr\vert \Omega \Bigr\rangle \Biggr] =
-\frac{\partial}{\partial z^{\nu}} \Bigl[a_{x'}^{D-1} a_z B\Bigl(y(x';z)
\Bigr)\Bigr] } \nonumber \\
& & \hspace{1.4cm} + \delta^0_{\nu} \Biggl\{ \frac{\partial}{\partial z^0}
\Bigl[a_{x'}^{D-1} a_z B\Bigl(y(x';z)\Bigr)\Bigr] +
\frac{\partial}{\partial x^{\prime 0}} \Bigl[a_{x'}^{D-1} a_z 
C\Bigl(y(x';z)\Bigr)\Bigr] \Biggr\} , \qquad \\
\lefteqn{\frac{\partial}{\partial z^{\prime \alpha}} \Biggl[ \sqrt{-g(z')}
g^{\alpha\beta}(z') \Bigl\langle \Omega \Bigl\vert T\Bigl[ A_{\mu}(x) 
A_{\beta}(z')\Bigr] \Bigr\vert \Omega \Bigr\rangle \Biggr] =
-\frac{\partial}{\partial x^{\mu}} \Bigl[a_x a_{z'}^{D-1} B\Bigl(y(x;z')
\Bigr)\Bigr] } \nonumber \\
& & \hspace{1.4cm} + \delta^0_{\mu} \Biggl\{ \frac{\partial}{\partial x^0}
\Bigl[a_x a_{z'}^{D-1} B\Bigl(y(x;z')\Bigr)\Bigr] +
\frac{\partial}{\partial z^{\prime 0}} \Bigl[a_x a_{z'}^{D-1}
C\Bigl(y(x;z')\Bigr)\Bigr] \Biggr\} , \qquad \\
\lefteqn{\frac{\partial}{\partial x^{\prime \rho}} 
\frac{\partial}{\partial z^{\prime \alpha}} \Biggl[ \! \sqrt{-g(x')} 
g^{\rho\sigma}(x') \sqrt{-g(z')} g^{\alpha\beta}(z')
\Bigl\langle \Omega \Bigl\vert T\Bigl[ A_{\sigma}(x') 
A_{\beta}(z')\Bigr] \Bigr\vert \Omega \Bigr\rangle \! \Biggr] =} \nonumber \\
& & \hspace{-.5cm} \frac{\partial}{\partial x^{\prime i}}
\frac{\partial}{\partial z^{\prime i}} \Bigl[ (a_{x'} a_{z'})^{D-1} B\Bigl(
y(x';z')\Bigr)\Bigr] - \frac{\partial}{\partial x^{\prime 0}}
\frac{\partial}{\partial z^{\prime 0}} \Bigl[ (a_{x'} a_{z'})^{D-1} C\Bigl(
y(x';z')\Bigr)\Bigr] \; . \qquad
\end{eqnarray}
All of this suggests that we would do well to organize the transformed
propagator into a doubly differentiated ``Integral Term'' and the
remaining, ``Other Term'',
\begin{equation}
\Bigl\langle \Omega \Bigl\vert T^*\Bigl[ \overline{A}_{\mu}(x)
\overline{A}_{\nu}(z)\Bigr] \Bigr\vert \Omega \Bigr\rangle =
\frac{\partial}{\partial x^{\mu}} \frac{\partial}{\partial z^{\nu}} 
\overline{\mathcal{I}}(x;z) + \Bigl[\mbox{}_{\mu} \overline{\mathcal{O}}_{\nu}
\Bigr](x;z) \; . \label{breakup}
\end{equation}
The Integral Term is,
\begin{eqnarray}
\lefteqn{ \overline{\mathcal{I}}(x;z) \equiv } \nonumber \\
& & \hspace{-.5cm} a_z \int_{V} \! d^Dx' \, G(x;x') a_{x'}^{D-1} 
B\Bigl( y(x';z)\Bigr) + a_x \int_{V} \! d^Dz' \, G(z;z') 
a_{z'}^{D-1} B\Bigl( y(x;z')\Bigr) \nonumber \\
& & \hspace{-.5cm} + \int_{V} \! d^Dx' \, G(x;x') \int_{V} \! d^Dz' \, G(z;z') 
\Biggl\{ \frac{\partial}{\partial x^{\prime i}}
\frac{\partial}{\partial z^{\prime i}} \Bigl[ (a_{x'} a_{z'})^{D-1} B\Bigl(
y(x';z')\Bigr)\Bigr] \nonumber \\
& & \hspace{5.4cm} - \frac{\partial}{\partial x^{\prime 0}}
\frac{\partial}{\partial z^{\prime 0}} \Bigl[ (a_{x'} a_{z'})^{D-1} C\Bigl(
y(x';z')\Bigr)\Bigr] \Biggr\} . \qquad
\end{eqnarray}
And the Other Term contains everything else,
\begin{eqnarray}
\lefteqn{ \Bigl[\mbox{}_{\mu} \overline{\mathcal{O}}_{\nu} \Bigr](x;z) \equiv
a_x a_z B\Bigl(y(x;z)\Bigr) \Bigl[\eta_{\mu\nu} \!+\! \delta^0_{\mu} 
\delta^0_{\nu}\Bigr] - a_x a_z C\Bigl(y(x;z)\Bigr) \delta^0_{\mu} 
\delta^0_{\nu} } \nonumber \\
& & - \delta^0_{\nu} \frac{\partial}{\partial x^{\mu}} 
\int_{V} \! d^Dx' \, G(x;x') \Biggl\{\frac{\partial}{\partial z^0}
\Bigl[a_{x'}^{D-1} a_z B\Bigl(y(x';z)\Bigr)\Bigr] \nonumber \\
& & \hspace{1cm} + \frac{\partial}{\partial x^{\prime 0}} \Bigl[a_{x'}^{D-1} 
a_z C\Bigl(y(x';z)\Bigr)\Bigr] \Biggr\} - \delta^0_{\mu} \frac{\partial}{
\partial z^{\nu}} \int_{V} \! d^Dz' \, G(z;z') \nonumber \\
& & \hspace{2.cm} \times \Biggl\{\! \frac{\partial}{\partial x^0}
\Bigl[a_x a_{z'}^{D-1} B\Bigl(y(x;z')\Bigr)\Bigr] 
+ \frac{\partial}{\partial z^{\prime 0}} \Bigl[a_x a_{z'}^{D-1} 
C\Bigl(y(x;z')\Bigr)\Bigr] \! \Biggr\} . \qquad
\end{eqnarray}

It remains to act the derivatives to simplify our expressions for $\overline{
\mathcal{I}}(x;z)$ and $[\mbox{}_{\mu} \overline{\mathcal{O}}_{\nu}](x;z)$. 
This is facilitated by some important identities obeyed by any function of 
$y(x;z)$,
\begin{eqnarray}
\lefteqn{\square_x F(y) = 
\frac{i 4 \pi^{\frac{D}2} \delta^D(x \!-\! z)}{\Gamma(\frac{D}2 \!-\! 1) \, 
H^{D-2} a^D} \times {\rm Res}[F] + H^2 \Biggl\{\! (4y \!-\! y^2) F'' + 
D (2 \!-\! y) F' \! \Biggr\} , \qquad } \label{FBox} \\
\lefteqn{\frac{\partial^2 F(y)}{\partial x^0 \partial z^0} =
\frac{i 4 \pi^{\frac{D}2} \delta^D(x \!-\! z)}{\Gamma(\frac{D}2 \!-\! 1) \, 
(H a)^{D-2}} \times {\rm Res}[F] + a_x a_z H^2 \Biggl\{ \Bigl[8 \!-\! (4 y 
\!-\! y^2) \Bigr] F'' } \nonumber \\
& & \hspace{3.9cm} - (2\!-\!y) F' + \Bigl(\frac{a_x}{a_z} \!+\! \frac{a_z}{a_x}
\Bigr) \Bigl[ -2(2 \!-\!y) F'' \!+\! 2 F'\Bigr] \Biggr\} 
\; , \qquad \label{F00} \\
\lefteqn{\frac{\partial^2 F(y)}{\partial x^i \partial z^i} = a_x a_z H^2
\Biggl\{4(2 \!-\!y) F'' -2 (D\!-\!1) F' - 4 \Bigl(\frac{a_x}{a_z} \!+\!
\frac{a_z}{a_x}\Bigr) F'' \Biggr\} \; , \qquad } \label{Fii} \\
\lefteqn{ H \Bigl[a_x \frac{\partial}{\partial z^0} \!+\! a_z \frac{\partial}{
\partial x^0}\Bigr] F(y) = a_x a_z H^2 \Biggl\{-2 (2 \!-\!y) F' + 
2 \Bigl(\frac{a_x}{a_z} \!+\! \frac{a_z}{a_x}\Bigr) F'\Biggr\} . \qquad }
\label{F0}
\end{eqnarray}
Here $\square$ is the covariant scalar d'Alembertian and ${\rm Res}[F]$ is 
the coefficient of $y^{1-\frac{D}2}$ in the Laurent expansion of $F(y)$. We 
shall also require some identities specific to $B(y)$ and $C(y)$,
\begin{eqnarray}
(4y \!-\! y^2) B''(y) + D(2 \!-\! y) B'(y) & = & (D \!-\! 2) B(y) 
\; , \label{BID}\\
(4y \!-\! y^2) C''(y) + D(2 \!-\! y) C'(y) & = & 2 (D \!-\! 3) C(y) 
\; . \label{CID}
\end{eqnarray}
And there is a very useful relation between $B(y)$ and $C(y)$,
\begin{equation}
C(y) = \frac12 (2\!-\!y) B(y) + \frac{k}{D\!-\!3} \qquad {\rm where} \qquad
k \equiv \frac{H^{D-2}}{(4\pi)^{\frac{D}2}} \frac{\Gamma(D\!-\!1)}{\Gamma(
\frac{D}2)} \; . \label{BCID}
\end{equation}
Note finally that substituting (\ref{BCID}) into (\ref{CID}) and using 
(\ref{BID}) implies,
\begin{equation}
(4y \!-\! y^2) B'(y) + (D\!-\!2) (2 \!-\! y) B(y) = -2 k \; . \label{BID2}
\end{equation}

It is best to start with the Other Term because it involves only first
derivatives. The reduction is straightforward for the $B$ term,
\begin{eqnarray}
\frac{\partial}{\partial z^0} \Bigl[ a_x^{D-1} a_z B\Bigl( y(x;z) \Bigr)\Bigr]
& = & a_x^{D-1} a_z \Bigl[ H a_z B + \frac{\partial y}{\partial z^0}
\, B'\Bigr] \; , \\
& = & a_x^D a_z H \Biggl\{ 2 B' + \frac{a_z}{a_x} \Bigl[- (2 \!-\!y) 
B' + B\Bigr] \Biggr\} \; . \qquad \label{Bcontr}
\end{eqnarray}
We begin the same way with the $C$ term but then use (\ref{BCID}) to convert
most of the $C$'s to $B$'s and simplify with (\ref{BID2}),
\begin{eqnarray}
\lefteqn{\frac{\partial}{\partial x^0} \Bigl[ a_x^{D-1} a_z C\Bigl( y(x;z) 
\Bigr)\Bigr] = a_x^{D-1} a_z \Bigl[ (D \!-\!1) H a_x C + \frac{\partial y}{
\partial x^0} \, C'\Bigr] \; , } \\
& & = a_x^D a_z H \Biggl\{ -(2 \!-\! y) C' + (D \!-\! 1) C + \frac{a_z}{a_x} 
\Bigl[2 C'\Bigr] \Biggr\} \; , \qquad \\
& & = a_x^D a_z H \Biggl\{ 2 C -\frac12 (2 \!-\! y)^2 B' + \frac12 (D \!-\! 2) 
(2 \!-\!y) B + k + \nonumber \\
& & \hspace{8.5cm} \frac{a_z}{a_x} \Bigl[(2 \!-\! y) B' \!-\! B\Bigr] \Biggr\} 
\; , \qquad \\
& & = a_x^D a_z H \Biggl\{ 2 C -2 B' + \frac{a_z}{a_x} \Bigl[(2 \!-\! y) B' 
\!-\! B\Bigr] \Biggr\} \; . \qquad \label{Ccontr}
\end{eqnarray}
Hence (\ref{Bcontr}) and (\ref{Ccontr}) almost completely cancel and our 
final result for the Other Term is,
\begin{eqnarray}
\lefteqn{ \Bigl[\mbox{}_{\mu} \overline{\mathcal{O}}_{\nu} \Bigr](x;z) \equiv
a_x a_z B\Bigl(y(x;z)\Bigr) \eta_{\mu\nu} + a_x a_z \Biggl\{ \frac12 y(x;z)
B\Bigl(y(x;z)\Bigr)\!-\! \frac{k}{D\!-\!3} \Biggr\} \delta^0_{\mu} 
\delta^0_{\nu} } \nonumber \\
& & \hspace{2cm} - 2 H a_z \delta^0_{\nu} \frac{\partial}{\partial x^{\mu}} 
\int_{V} \! d^Dx' \sqrt{-g(x')} \, G(x;x') C\Bigl(y(x';z)\Bigr) \nonumber \\
& & \hspace{3.3cm} - 2 H a_x \delta^0_{\mu} \frac{\partial}{\partial z^{\nu}} 
\int_{V} \! d^Dz' \sqrt{-g(z')} \, G(z;z') C\Bigl(y(x;z')\Bigr) . 
\qquad \label{Other}
\end{eqnarray}

The Integral term $\overline{\mathcal{I}}(x;z)$ involves second derivatives. 
We only need (\ref{Fii}) to reduce the spatial case,
\begin{eqnarray}
\lefteqn{\frac{\partial}{\partial x^i} \frac{\partial}{\partial z^i} 
\Bigl[ (a_{x} a_{z})^{D-1} B\Bigl(y(x;z)\Bigr)\Bigr] } \nonumber \\
& & \hspace{1cm} = H^2 (a_x a_z)^D \Biggl\{4 (2 \!-\!y)
B'' - 2(D\!-\!1) B' + \Bigl(\frac{a_x}{a_z} \!+\! \frac{a_z}{a_x}\Bigr)
\Bigl[ -4 B''\Bigr]\Biggr\} . \qquad
\end{eqnarray}
Reducing the temporal derivative term is more involved. We begin by passing
the derivatives through the scale factors, then employ relations (\ref{F00})
and (\ref{F0}), and convert most of the $C(y)$'s to $B(y)$ using (\ref{BCID}),
eliminating second derivatives with (\ref{BID}-\ref{CID}) as needed. The
result is,
\begin{eqnarray}
\lefteqn{\frac{\partial}{\partial x^0} \frac{\partial}{\partial z^0} 
\Bigl[ (a_x a_z)^{D-1} C\Bigl( y(x;z)\Bigr)\Bigr] = (a_x a_z)^{D-1} 
\Biggl\{ \frac{\partial}{\partial x^0} \frac{\partial}{\partial z^0} }
\nonumber \\
& & \hspace{1.5cm} + (D\!-\!1) H \Bigl[a_x \frac{\partial}{\partial z^0} \!+\!
a_z \frac{\partial}{\partial x^0}\Bigr] + (D\!-\!1)^2 a_x a_z H^2\Biggr\} 
C\Bigl(y(x;z)\Bigr) \; , \qquad \\
& & = i a_x^D \delta^D(x\!-\! z') + H^2 (a_x a_z)^D \Biggl\{4 C + 4(2\!-\!y)
B'' - 2 (D\!+\!3) B' \nonumber \\
& & \hspace{4.9cm} + \Bigl(\frac{a_x}{a_z} \!+\! \frac{a_z}{a_x}\Bigr)
\Bigl[-4 B'' + 2 (2 \!-\! y) B' - 2 B\Bigr] \Biggr\} . \qquad
\end{eqnarray}
Adding the two terms gives a compact form,
\begin{eqnarray}
\lefteqn{\frac{\partial}{\partial x^i} \frac{\partial}{\partial z^i} 
\Bigl[ (a_{x} a_{z})^{D-1} B(y)\Bigr] - \frac{\partial}{\partial x^0} 
\frac{\partial}{\partial z^0} \Bigl[ (a_x a_z)^{D-1} C(y)\Bigr] = 
-i a_x^D \delta^D(x \!-\! z) } \nonumber \\
& & \hspace{1.5cm} + (a_x a_z)^D H^2 \Biggl\{\! -4 C + 8 B'
+ \Bigl(\frac{a_x}{a_z} \!+\! \frac{a_z}{a_x}\Bigr) \Bigl[-2 (2 \!-\!y) B'
+ 2 B\Bigr] \! \Biggr\} . \qquad 
\end{eqnarray}
A further simplification can be effected by means of the identity,
\begin{equation}
(a_x a_z)^D \square_x \Bigl[ \frac{a_x}{a_z} B(y)\Bigr] = i a_x^D \delta^D(x 
\!-\! z) + (a_x a_z)^D H^2 \Biggl\{-4 B' + \frac{a_x}{a_z} \Bigl[2 (2 \!-\!y)
B' - 2 B\Bigr] \Biggr\} .
\end{equation}
Using this and the result with $x^{\mu}$ and $z^{\mu}$ interchanged gives,
\begin{eqnarray}
\lefteqn{\frac{\partial}{\partial x^i} \frac{\partial}{\partial z^i} 
\Bigl[ (a_{x} a_{z})^{D-1} B(y)\Bigr] - \frac{\partial}{\partial x^0} 
\frac{\partial}{\partial z^0} \Bigl[ (a_x a_z)^{D-1} C(y)\Bigr] = 
i a_x^D \delta^D(x \!-\! z) } \nonumber \\
& & \hspace{3.5cm} + (a_x a_z)^D \Biggl\{\! -4 H^2 C - \square_x \Bigl[
\frac{a_x}{a_z} B\Bigr] - \square_z \Bigl[ \frac{a_z}{a_x} B\Bigr] 
\Biggr\} . \qquad
\end{eqnarray}

Our final result for the Integral Term involves the surface integral,
\begin{eqnarray}
\lefteqn{\mathcal{S}_B(x;z) \equiv \frac{a_x}{a_z} B\Bigl(y(x;z)\Bigr) 
\!-\! \int_{V} \!\! d^Dx' \sqrt{-g(x')} \, G(x;x') \square_{x'} \Biggl[\!
\frac{a_{x'}}{a_z} B\Bigl(y(x';z)\Bigr) \!\Biggr] , \qquad } \\
& & \hspace{-.7cm} = \!\! \int_{\partial V} \!\!\!\!\!\! d^{D-1}\!x_{\mu}' \!
\sqrt{-g'} \!g^{\prime \mu\nu} \! \Biggl[\! \frac{a_{x'}}{a_z} i\Delta_B(x';z) 
\partial_{\nu}' G(x;x') \!-\! G(x;x') \partial_{\nu}' \Bigl[ 
\frac{a_{x'}}{a_z} i\Delta_B(x';z)\Bigr] \! \Biggr] . \qquad 
\end{eqnarray}
This function is obviously homogeneous; that is, $\square_x$ annihilates it. 
In the next section we will show how to choose the homogeneous contribution 
to the full gauge parameter $\theta[A](x)$ so as to cancel it and similar
terms. The final result for the Integral Term is,
\begin{eqnarray}
\lefteqn{\overline{\mathcal{I}}(x;z) = \int_{V} \! d^Dx' \sqrt{-g(x')} \, 
G(x;x') \mathcal{S}_B(z;x') } \nonumber \\
& & + \int_{V} \! d^Dz' \sqrt{-g(z')} \, G(z;z') \mathcal{S}_B(x;z') 
+ i \int_{V} \! d^Dx' \sqrt{-g(x')} \, G(x;x') G(z;x') \nonumber \\
& & - 4 H^2 \int_{V} \! d^Dx' \sqrt{-g(x')} \, G(x;x') 
\int_{V} \! d^Dz' \sqrt{-g(z')} \, G(z;z') C\Bigl( y(x';z')\Bigr) \; . 
\qquad \label{Integral}
\end{eqnarray}

\section{The Invariant Propagator}

The purpose of this section is to show that the transformed propagator of
the previous section agrees, up to surface terms, with the unique de Sitter 
invariant propagator which was found by solving the propagator equation in 
Lorentz gauge \cite{TW7}. Of course we begin by describing that solution.
We then decompose it in analogy with the scheme (\ref{breakup}) of the 
previous section, into an ``Integral Term'' and an ``Other Term.'' At this
stage there is a digression to derive an identity for the convolution of 
scalar propagators. The section closes by applying this identity to 
demonstrate that the two propagators agree up to surface integrals.

The Lorentz gauge propagator equation has a unique de Sitter invariant
solution which can expressed in terms of a function $\gamma(y)$ \cite{TW7},
\begin{eqnarray}
\lefteqn{ i\Bigl[\mbox{}_{\mu} \Delta^{\rm dS}_{\nu}\Bigr](x;z) =
\frac1{4 (D\!-\!1) H^2} \Biggl\{ \frac{\partial^2 y(x;z)}{\partial x^{\mu}
\partial z^{\nu}} \Bigl[-(4y \!-\! y^2) \gamma' \!-\! (D\!-\!1)
(2\!-\!y) \gamma\Bigr] } \nonumber \\
& & \hspace{6cm} + \frac{\partial y}{\partial x^{\mu}} \,
\frac{\partial y}{\partial z^{\nu}} \Bigl[ (2 \!-\!y) \gamma'
\!-\! (D\!-\!1) \gamma\Bigr] \! \Biggr\} . \qquad \label{dSprop}
\end{eqnarray}
The function $\gamma(y)$ has a very complicated series expansion,
\begin{eqnarray}
\lefteqn{\gamma(y) = \frac12 \Bigl(\frac{D\!-\!1}{D\!-\!3}\Bigr)
\frac{H^{D-2}}{(4\pi)^{\frac{D}2}} \Biggl\{ (D\!-\!3) \Gamma\Bigl(\frac{D}2
\!-\!1\Bigr) \Bigl(\frac{4}{y}\Bigr)^{\frac{D}2-1} \!\!+ \sum_{n=0}^{\infty}
\Biggl[ \frac{(n\!+\!1) \Gamma(n \!+\!D \!-\!1)}{\Gamma(n \!+\! \frac{D}2
\!+\!1)} } \nonumber \\
& & \hspace{2.5cm} \times \Bigl[\psi\Bigl(2\!-\!\frac{D}2\Bigr) \!-\!
\psi\Bigl(\frac{D}2 \!-\! 1\Bigr) \!+\! \psi(n \!+\! D \!-\! 1) \!-\!
\psi(n \!+\! 2)\Bigr] \Bigl(\frac{y}{4}\Bigr)^n \nonumber \\
& & - \frac{(n\!-\!\frac{D}2 \!+\! 3) \Gamma(n \!+\! \frac{D}2 \!+\!1)}{
\Gamma(n \!+\! 3)} \Bigl[\psi\Bigl(2\!-\!\frac{D}2\Bigr) \!-\!
\psi\Bigl(\frac{D}2 \!-\! 1\Bigr) \nonumber \\
& & \hspace{4cm} + \psi\Bigl(n \!+\! \frac{D}2 \!+\! 1\Bigr) \!-\!
\psi\Bigl(n \!-\! \frac{D}2 \!+\! 4\Bigr)\Bigr] \Bigl(\frac{y}{4}\Bigr)^{n
-\frac{D}2 + 2} \Biggr] \Biggr\} . \qquad \label{gamma}
\end{eqnarray}
Although it might seem unwieldy, this formalism has been used to perform
several two loop computations in scalar quantum electrodynamics 
\cite{PTW1,PTW2}.

The function $\gamma(y)$ obeys the second order differential equation,
\begin{equation}
(4y \!-\! y^2) \gamma'' + (D\!+\!2) (2\!-\!y) \gamma' - 2 (D\!-\!1) \gamma
= 2 (D\!-\!1) B'(y) \; . \qquad \label{gameqn}
\end{equation}
One consequence is,
\begin{equation}
\frac{\partial}{\partial y} \Biggl[-(4y \!-\! y^2) \gamma' \!-\! (D\!-\!1)
(2 \!-\!y) \gamma \!+\! 2(D\!-\!1) B\Biggr] = (2 \!-\! y) \gamma' \!-\! 
(D\!-\!1) \gamma \; .
\end{equation}
Hence we can decompose the invariant propagator in a form analogous to
that of the transformed propagator (\ref{breakup}),
\begin{eqnarray}
\lefteqn{ i\Bigl[\mbox{}_{\mu} \Delta^{\rm dS}_{\nu}\Bigr](x;z) = -
\frac1{2 H^2} B\Bigl(y(x;z)\Bigr) \, \frac{\partial^2 y(x;z)}{\partial x^{\mu}
\partial z^{\nu}} } \nonumber \\
& & \hspace{-.1cm} + \frac1{4 (D\!-\! 1) H^2} \frac{\partial}{\partial x^{\mu}}
\frac{\partial}{\partial z^{\nu}} \, I\Biggl[-(4y \!-\! y^2) \gamma' \!-\!
(D\!-\!1) (2 \!-\!y) \gamma \!+\! 2(D\!-\!1) B\Biggr] , \qquad \label{invprop}
\end{eqnarray}
where the notation ``$I[f]$'' of a function $f(y)$ stands for its indefinite
integral,
\begin{equation}
I[f](y) \equiv \int^y \!\! dy' \, f(y') \; .
\end{equation}

At this point it is useful to digress on the subject of scalar propagators.
Three were introduced in the previous section --- $+i \times G(x;z)$, 
$i\Delta_B(x;z)$ and $i\Delta_C(x;z)$ --- and it might seem that there is a 
bewildering variety of them, each with its own important special properties. 
However, a unified treatment can be given in terms of the equation,
\begin{equation}
\sqrt{-g(x)} \, \Biggl\{ \square_x + \Bigl[ \nu^2 \!-\! \Bigl(\frac{D-1}2
\Bigr)^2 \Bigr] H^2 \Biggr\} i\Delta_{\nu}(x;z) = i \delta^D(x \!-\! z) \; .
\label{nuprop}
\end{equation}
The three propagators of the previous section correspond to the following
choices for $\nu$:
\begin{eqnarray}
i \times G(x;z) & \Longrightarrow & \nu = \Bigl(\frac{D-1}2\Bigr) \; , \\
i \Delta_B(x;z) & \Longrightarrow & \nu = \Bigl(\frac{D-3}2\Bigr) \; , \\
i \Delta_C(x;z) & \Longrightarrow & \nu = \Bigl(\frac{D-5}2\Bigr) \; .
\end{eqnarray}
For general $\nu$ the spatial plane wave mode functions corresponding to
Bunch-Davies vacuum are,
\begin{equation}
u_{\nu}(x^0,k) \equiv \sqrt{\frac{\pi}{4 H}} \; a^{-\frac{D-1}2} \, 
H^{(1)}_{\nu}(-k x^0) \; . \label{unu}
\end{equation}
When it exists, the Fourier mode sum for the propagator is \cite{JMPW},
\begin{eqnarray}
\lefteqn{i\Delta_{\nu}(x;z) = \int \!\! \frac{d^{D-1}k}{(2\pi)^{D-1}} \, 
e^{i \vec{k} \cdot (\vec{x} - \vec{z})} \Biggl\{ \theta(x^0 \!-\! z^0) 
u_{\nu}(x^0,k) u^*_{\nu}(z^0,k) } \nonumber \\
& & \hspace{6.4cm} + \theta(z^0 \!-\! x^0) u^*_{\nu}(x^0,k) u_{\nu}(z^0,k)
\Biggr\} . \qquad \label{modesum}
\end{eqnarray}
When this sum exists the result is de Sitter invariant \cite{CT},
\begin{equation}
i\Delta_{\nu}(x;z) = \frac{H^{D-2}}{(4\pi)^{\frac{D}2}} 
\frac{\Gamma(\frac{D-1}2 \!+\! \nu) \Gamma(\frac{D-1}2 \!-\! \nu)}{
\Gamma(\frac{D}2)} \, \mbox{}_2 F_1\Bigl( \frac{D-1}2 \!+\! \nu,
\frac{D-1}2 \!-\! \nu; \frac{D}2;1 \!-\! \frac{y}4\Bigr) \; .
\end{equation}

When the Fourier mode sum (\ref{modesum}) is infrared divergent it must be 
cut off, either by making the mode functions less singular for super-horizon 
wave lengths \cite{Vilenkin} or by working on a spatially compact manifold 
\cite{TW8}. Either procedure breaks de Sitter invariance. A special case of 
some importance to our discussion is $\nu = (D-1)/2$, for which the result is
\cite{OW,JMPW},
\begin{equation}
\nu = \Bigl(\frac{D\!-\!1}2\Bigr) \qquad \Longrightarrow \qquad 
i\Delta_A(x;z) = A\Bigl( y(x;z)\Bigr) + k \ln(a_x a_z) \; , \label{DeltaA}
\end{equation}
where the constant $k$ was defined in (\ref{BCID}) and the function $A(y)$
has the expansion,
\begin{eqnarray}
\lefteqn{A(y) \equiv \frac{H^{D-2}}{(4\pi)^{\frac{D}2}} \Biggl\{
\frac{\Gamma(\frac{D}2)}{\frac{D}2 \!-\! 1}
\Bigl(\frac{4}{y}\Bigr)^{ \frac{D}2 -1} \!+\!
\frac{\Gamma(\frac{D}2 \!+\! 1)}{\frac{D}2 \!-\! 2}
\Bigl(\frac{4}{y} \Bigr)^{\frac{D}2-2} \!-\! \pi
\cot\Bigl(\frac{\pi D}2\Bigr)
\frac{\Gamma(D \!-\! 1)}{\Gamma(\frac{D}2)} } \nonumber \\
& & \hspace{.7cm} + \sum_{n=1}^{\infty} \Biggl[\frac1{n}
\frac{\Gamma(n \!+\! D \!-\! 1)}{\Gamma(n \!+\! \frac{D}2)}
\Bigl(\frac{y}4 \Bigr)^n \!\!\!\! - \frac1{n \!-\! \frac{D}2 \!+\!
2} \frac{\Gamma(n \!+\!  \frac{D}2 \!+\! 1)}{ \Gamma(n \!+\! 2)}
\Bigl(\frac{y}4 \Bigr)^{n - \frac{D}2 +2} \Biggr] \Biggr\} . \qquad
\end{eqnarray}
As with the expansions (\ref{Bexp}-\ref{Cexp}) for $B(y)$ and $C(y)$, the
infinite series terms of $A(y)$ vanish for $D=4$, so they only need to be
retained when multiplying a potentially divergent quantity, and even then
one only needs to include a handful of them. This makes loop computations
manageable. For a massless, minimally coupled scalar with a quartic 
self-interaction, two loop results have been obtained for the expectation 
value of the stress tensor \cite{OW}, for the scalar self-mass-squared
\cite{BOW} and for the quantum-corrected mode functions \cite{KO}. In
Yukawa theory it has been used to compute the expectation vlaue of the
coincident vertex function at two loop order \cite{MW2}, and it has been
used for a variety of two loop computations in scalar quantum electrodynamics
\cite{PTW1,PTW2}. It should also be noted that the de Sitter breaking 
correction to $i\Delta_A(x;z)$ in expression (\ref{DeltaA}) can be derived 
from the infrared-truncated mode sum \cite{JMPW}, and it serves to reproduce 
the classic and well known result for the coincidence limit of the propagator 
\cite{classic}.

The function $A(y)$ obeys a differential equation analogous to (\ref{BID})
and (\ref{CID}),
\begin{equation}
(4 y \!-\! y^2) A'' + D (2 \!-\! y) A' = (D \!-\! 1) k \; . \label{AID}
\end{equation}
A number of identities relate the derivative of $A(y)$ to $B(y)$ and $C(y)$,
\begin{eqnarray}
A' & = & -\frac12 (D \!-\!3) B + C' \; , \label{ABC} \\
(4 y \!- y^2) A' & = & -2 (D \!-\! 2) B - (2 \!-\! y) k \; .
\end{eqnarray}
It is also useful to note the result of acting the scalar d'Alembertian on
a function of the scale factor,
\begin{equation}
\square f(a) = -H^2 \Bigl[a^2 f''(a) + D a f'(a)\Bigr] \; . \label{aID}
\end{equation}

Now consider Green's second identity for any two functions $F(x')$ and 
$G(x')$,
\begin{eqnarray}
\lefteqn{F(x') \sqrt{-g(x')} \, \square' G(x') - G(x') \sqrt{-g(x')} \,
\square' F(x')} \nonumber \\
& & \hspace{2cm} = \partial_{\mu}' \Biggl\{ \sqrt{-g(x')} \, g^{\mu\nu}(x') 
\Bigl[ F(x') \partial_{\nu}' G(x') - G(x') \partial_{\nu}' F(x')\Bigr] 
\Biggr\} . \qquad \label{Green2}
\end{eqnarray}
We choose $G(x')$ to be any symmetric Green's function $G(x;x') = G(x';x)$,
\begin{equation}
\sqrt{-g(x)} \, \square G(x;x') = \delta^D(x \!-\! x') \; .
\end{equation}
We can obviously integrate (\ref{Green2}) over any region $V$ with boundary
$\partial V$ to conclude,
\begin{eqnarray}
\lefteqn{ F(x) = \int_{V} \!\! d^Dx' \sqrt{-g(x')} \, G(x;x') \square_{x'} 
F(x') } \nonumber \\
& & \hspace{1cm} + \int_{\partial V} \!\! d^{D-1}\!x_{\mu}' \sqrt{-g'} \,
g^{\prime\mu\nu} \Biggl[F(x') \partial_{\nu}' G(x;x') - G(x;x') \partial_{\nu}'
F(x') \Biggr] \; . \qquad \label{intrel}
\end{eqnarray}
Relation (\ref{intrel}) is true for any Green's function so we are free to 
use the $A$-type propagator, $G(x;x') = -i \times i\Delta_A(x;x')$. Relation
(\ref{intrel}) is also valid for any function $F(x)$ so we are free to
make the choice,
\begin{equation}
F(x) \longrightarrow \frac{i\Delta_{\nu}(x;z) \!-\! i\Delta_A(x;z)}{[
(\frac{D-1}2)^2 \!-\! \nu^2] \, H^2} \; .
\end{equation}
The surface terms involving $i\Delta_A$ obviously cancel so the result is,
\begin{eqnarray}
\lefteqn{-i \!\! \int_{V} \! d^Dx' \sqrt{-g(x')} \, i\Delta_A(x;x') 
i\Delta_{\nu}(x';z) = \frac{i\Delta_{\nu}(x;z) \!-\! i\Delta_A(x;z)}{[(
\frac{D-1}2)^2 \!-\! \nu^2] H^2} } \nonumber \\
& & \hspace{-.7cm} +i \!\! \int_{\partial V} \!\!\!\! d^{D-1}\!x'_{\rho} 
\sqrt{-g'} \, g^{\prime \rho\sigma} \Biggl[\frac{i\Delta_{\nu}(x';z) 
\partial_{\sigma}' i\Delta_A(x;x')\!-\!i\Delta_A(x;x') \partial_{\sigma}'
i\Delta_{\nu}(x';z)}{[(\frac{D-1}2)^2 \!-\! \nu^2] H^2} \Biggr] \! . \qquad 
\label{convolution}
\end{eqnarray}
We call (\ref{convolution}) the ``Convolution Identity.''

Choosing $\nu = (D-5)/2$ in the Convolution Identity (\ref{convolution})
gives us a relation for the $C$-type propagator $C(y)$,
\begin{equation}
-i \!\! \int_{V} \! d^Dx' \sqrt{-g(x')} \, i\Delta_A(x;x') 
i\Delta_C(x';z) = \frac{i\Delta_C(x;z) \!-\! i\Delta_A(x;z)
\!-\! \mathcal{S}_C(x;z)}{2 (D \!-\! 3) H^2} \; , \label{convC}
\end{equation}
where the surface term is,
\begin{equation}
\mathcal{S}_C(x;z) \equiv \!\! \int_{\partial V}\!\!\!\!\!\! d^{D-1}\!x'_{\rho} 
\!\sqrt{-g'} g^{\prime \rho\sigma} \Bigl[i\Delta_C(x';z) 
\partial_{\sigma}' G(x;x')\!-\! G(x;x') \partial_{\sigma}'
i\Delta_C(x';z) \Bigr] . \label{surfC}
\end{equation}
We now substitute (\ref{convC}) in our result (\ref{Other}) for the the Other
Term of the previous section,
\begin{eqnarray}
\lefteqn{ \Bigl[\mbox{}_{\mu} \overline{\mathcal{O}}_{\nu} \Bigr](x;z) =
a_x a_z B\Bigl(y(x;z)\Bigr) \eta_{\mu\nu} + a_x a_z \Biggl\{ \frac12 y(x;z)
B\Bigl(y(x;z)\Bigr)\!-\! \frac{k}{D\!-\!3} \Biggr\} \delta^0_{\mu} 
\delta^0_{\nu} } \nonumber \\
& & \hspace{3cm} - 2 H \Biggl[a_z \delta^0_{\nu} \frac{\partial}{\partial 
x^{\mu}} \!+\! a_x \delta^0_{\mu} \frac{\partial}{\partial z^{\nu}} \Biggr]
\Biggl\{ \frac{i\Delta_C(x;z) \!-\! i\Delta_A(x;z)}{2 (D \!-\! 3) H^2} 
\Biggr\} \nonumber \\
& & \hspace{4cm} + \frac{a_z \delta^0_{\nu}}{(D \!-\! 3) H} \frac{\partial 
\mathcal{S}_C(x;z)}{\partial x^{\mu}} + \frac{a_x \delta^0_{\mu}}{(D \!-\! 3)H}
\frac{\partial \mathcal{S}_C(z;x)}{\partial z^{\nu}} \; . \qquad
\end{eqnarray}
The derivative is easy to simplify using (\ref{ABC}),
\begin{eqnarray}
\lefteqn{- 2 H a_z \delta^0_{\nu} \frac{\partial}{\partial x^{\mu}} \Biggl\{ 
\frac{i\Delta_C(x;z) \!-\! i\Delta_A(x;z)}{2 (D \!-\! 3) H^2}\Biggr\} } 
\nonumber \\
& & \hspace{3cm} = -\frac{a_z \delta^0_{\nu}}{(D \!-\!3) H} \, 
\frac{\partial y}{\partial x^{\mu}} \, \Bigl(C' - A'\Bigr) + \frac{k}{D\!-\!3}
\, \delta^0_{\mu} \delta^0_{\nu} \, a_x a_z \; , \qquad \\
& & \hspace{3cm} = a_x a_z \Biggl\{ -a_z H \Delta x_{\mu} \delta^0_{\nu} B
-\frac12 y B \delta^0_{\mu} \delta^0_{\nu} + \frac{k}{D \!-\! 3} \, 
\delta^0_{\mu} \delta^0_{\nu} \Biggr\} . \qquad
\end{eqnarray}
Combining everything results in an expression for the Other Term which is
almost de Sitter invariant, modulo the surface terms,
\begin{eqnarray}
\lefteqn{ \Bigl[\mbox{}_{\mu} \overline{\mathcal{O}}_{\nu} \Bigr](x;z) =
a_x a_z \Biggl\{ -\frac12 y \delta^0_{\mu} \delta^0_{\nu} - a_z H \Delta 
x_{\mu} \delta^0_{\nu} + a_x \delta^0_{\mu} H \Delta x_{\nu} + \eta_{\mu\nu}
\Biggr\} B } \nonumber \\
& & \hspace{.5cm} + \frac{k}{D \!-\!3} \, a_x a_z \delta^0_{\mu} \delta^0_{\nu}
+ \frac{a_z \delta^0_{\nu}}{(D \!-\! 3) H} \frac{\partial \mathcal{S}_C(x;z)}{
\partial x^{\mu}} + \frac{a_x \delta^0_{\mu}}{(D \!-\!3) H} \frac{\partial 
\mathcal{S}_C(z;x)}{\partial z^{\nu}} \; , \qquad \\
& & \hspace{0cm}=\!-\frac{\partial^2 y(x;z)}{\partial x^{\mu}\partial z^{\nu}}
\frac{B}{2 H^2} \!+\! \frac{k a_x a_z \delta^0_{\mu} \delta^0_{\nu}}{D 
\!-\!3} \!+\! \frac{a_z \delta^0_{\nu}}{D \!-\!3} \frac{\partial 
\mathcal{S}_C(x;z)}{\partial H x^{\mu}} \!+\! \frac{a_x \delta^0_{\mu}}{
D \!-\! 3} \frac{\partial \mathcal{S}_C(z;x)}{\partial H z^{\nu}} . 
\qquad \label{Obar}
\end{eqnarray}

The first term on the right hand side of (\ref{Obar}) agrees with the first
term in our decomposition (\ref{invprop}) for the invariant propagator. We
must obviously choose the homogeneous contributions to the gauge parameter
$\theta[A](x)$ so as to cancel the surface terms in (\ref{Obar}). That leaves 
the term proportional to $k$, which relation (\ref{aID}) allows us to 
recognize as a potential part of the Integral Term,
\begin{equation}
\frac{k a_x a_z \delta^0_{\mu} \delta^0_{\nu}}{D \!-\!3} = \frac{\partial}{
\partial x^{\mu}} \, \frac{\partial}{\partial z^{\nu}} \Biggl\{
\frac{k \ln^2(a_x a_z)}{2 (D \!-\! 3) H^2}  + {\rm const} \times \ln(a_x a_z) 
\Biggr\} \; . \label{logsq}
\end{equation}
We will presently see that precisely the bracketed expression is needed to 
make $\overline{\mathcal{I}}(x;z)$ de Sitter invariant, up to surface terms.

A matter of great importance for us is what the Convolution Identity 
(\ref{convolution}) gives when the index $\nu$ is chosen to be $(D\!-\!1)/2$,
corresponding to the $A$-type propagator. The term on the right hand side
obviously gives a derivative with respect to the index $\nu$,
\begin{eqnarray}
\lim_{\nu \rightarrow (\frac{D-1}2)} \Biggl[ \frac{i\Delta_{\nu}(x;z) \!-\! 
i\Delta_A(x;z)}{[(\frac{D-1}2)^2 \!-\! \nu^2] H^2} \Biggr] & = & -
\frac{\frac{\partial}{\partial \nu} \, i\Delta_{\nu}(x;z) }{(D \!-\!1) H^2}
\Biggl\vert_{\nu = (\frac{D-1}2)} \; , \qquad \\
& \equiv & -\frac{i\Delta_{A'}(x;z)}{(D \!-\!1) H^2} \; . \label{DeltaA'}
\end{eqnarray}
Hence the convolution of two $A$-type propagators gives,
\begin{equation}
-i \!\! \int_{V} \! d^Dx' \sqrt{-g(x')} \, i\Delta_A(x;x') i\Delta_A(x';z) 
= -\frac{i\Delta_{A'}(x;z)}{(D \!-\!1) H^2} + \frac{\mathcal{S}_A(x;z)}{
(D \!-\!1) H^2} \; ,
\end{equation}
where the surface term is,
\begin{equation}
\mathcal{S}_A(x;z) \!\equiv \!\!
\int_{\partial V} \!\!\!\!\!\! d^{D-1}\!x'_{\rho} \! \sqrt{-g'} 
g^{\prime \rho\sigma} \Bigl[i\Delta_{A'}(x';z) \partial_{\sigma}' G(x;x')
\!-\! G(x;x') \partial_{\sigma}' i\Delta_{A'}(x';z) \Bigr] .
\end{equation}

Like the $A$ propagator, the $A'$ propagator breaks de Sitter invariance.
The simplest way to see this is by differentiating relation (\ref{nuprop})
with respect to $\nu$ and then setting $\nu = (D-1)/2$,
\begin{eqnarray}
0 & = & \frac{\partial}{\partial \nu} \Biggl\{ \Biggl[\square_x + \Bigl[\nu^2 -
\Bigl(\frac{D\!-\!1}2\Bigr)^2\Bigr] H^2 \Biggr] i\Delta_{\nu}(x;z) 
\Biggr\}_{\nu = (\frac{D-1}2)} \; , \qquad \\
& = & \square_x i\Delta_{A'}(x;z) + (D\!-\!1) H^2 i\Delta_A(x;z) \; . 
\qquad \label{BoxA}
\end{eqnarray}
Because $i\Delta_A(x;z) = A(y) + k \ln(a_x a_z)$ has a de Sitter breaking
term it is clear that $i\Delta_{A'}(x;z)$ must as well. From relation 
(\ref{aID}) we infer,
\begin{equation}
i\Delta_{A'}(x;z) = \mathcal{A}\Bigl(y(x;z)\Bigr) + {\rm const} \times
\ln(a_x a_z) + \frac12 k \ln^2(a_x a_z) \; . \label{A'form}
\end{equation}
We do not require the coefficient of the $\ln(a_x a_z)$ term but the series
expansion for $\mathcal{A}(y)$ is,
\begin{eqnarray}
\lefteqn{\mathcal{A}(y) = \frac{H^{D-2}}{(4\pi)^{\frac{D}2}} \, 
\Bigl(\frac{4}{y}\Bigr)^{\frac{D}2 - 2} \!\!\times 2 \Bigl(\frac{D \!-\! 1}{D 
\!-\! 4}\Bigr) \Gamma\Bigl(\frac{D}2 \!-\! 1\Bigr) - \frac{H^{D-2}}{(4\pi)^{
\frac{D}2}} \sum_{n=0}^{\infty} \Biggl\{ \frac{(\frac14 y)^{n -\frac{D}2 +3}}{
n \!-\! \frac{D}2 \!+\!3} } \nonumber \\
& & \hspace{0cm} \times \frac{\Gamma(n\!+\!\frac{D}2 \!+\! 2)}{(n \!+\! 2)!} 
\Biggl[\psi\Bigl(2 \!-\! \frac{D}2\Bigr) \!-\! \psi\Bigl(\frac{D}2 \!-\! 1
\Bigr) \!+\! \psi\Bigl(n \!+\! \frac{D}2 \!+\! 2\Bigr) \!-\! \psi\Bigl(n \!-\!
\frac{D}2 \!+\! 3\Bigr)\Biggr] \nonumber \\
& & \hspace{-.7cm} - \frac{(\frac14 y)^{n+1}}{n \!+\! 1} \!\times\!
\frac{\Gamma(n \!+\! D)}{\Gamma(n \!+\! \frac{D}2 \!+\! 1)} 
\Biggl[ \psi\Bigl(2 \!-\! \frac{D}2 \Bigr) \!-\! \psi\Bigl(\frac{D}2 \!-\! 1
\Bigr) \!+\! \psi(n \!+\! D) \!-\! \psi(n \!+\! 1)\Biggr] \!\Biggr\} . \qquad
\label{calA}
\end{eqnarray}

One can hardly fail to notice the similarities in the series expansion
(\ref{gamma}) for $\gamma(y)$ and the expansion (\ref{calA}) we have just 
given for the de Sitter invariant part of the $A'$ propagator. The relation
between them is,
\begin{equation}
\mathcal{A}(y) = \frac14 \Bigl(\frac{D \!-\! 3}{D \!-\! 1}\Bigr) 
I\Bigl[(4 y \!-\! y^2) \gamma' + (D\!-\!1) (2 \!-\! y) \gamma\Bigr] 
-\frac12 (D \!-\!2) I[B] + {\rm const} \; . \label{calAgam}
\end{equation}
It is tedious but straightforward to check (\ref{calAgam}) using the
series expansions but a simpler way of recognizing it is to act the scalar
d'Alembertian on both sides. In view of (\ref{BoxA}) the left hand side gives,
\begin{equation}
\frac{\square}{H^2} \mathcal{A}(y) = -(D\!-\! 1) A(y) + {\rm const} \; .
\end{equation}
To compute the right hand side we need the lemma,
\begin{equation}
(2\!-\!y) (4y \!-\! y^2) \gamma' + (4y \!-\! y^2) \gamma + D (2 \!-\! y)^2 
\gamma = 2 (D \!-\! 1) I\Bigl[ (2 \!-\! y) B'\Bigr] + {\rm const} \; .
\label{lemma}
\end{equation}
This follows from differentiation with respect to $y$ and using the
equation (\ref{gameqn}) for $\gamma(y)$. Now act $\square/H^2$ on the first
term on the right hand side of (\ref{calAgam}), then use the $\gamma$ 
equation (\ref{gameqn}), and finally relations (\ref{BID2}) and (\ref{lemma}),
\begin{eqnarray}
\lefteqn{\frac{\square}{H^2} I\Bigl[(4 y \!-\! y^2) \gamma' \!+\! (D\!-\!1)
(2 \!-\!y) \gamma\Bigr] = (4y \!-\!y^2) \Bigl[(4y \!-\! y^2) \gamma'' \!+\! 
(D\!+\!1) (2 \!-\! y) \gamma' } \nonumber \\
& & \hspace{2.5cm} - (D\!-\!1) \gamma\Bigr] + D (2 \!-\!y) 
\Bigl[(4 y \!-\! y^2) \gamma' \!+\! (D \!-\!1) (2 \!-\!y) \gamma\Bigr] 
\; , \qquad \\
& & \hspace{-.3cm} = (D \!-\!1) \Biggl\{(2 \!-\! y) (4 y \!-\! y^2) \gamma' 
\!+\! (4 y \!-\! y^2) \gamma \!+\! D (2 \!-\! y)^2 \gamma \!+\! 2 (4 y \!-\! 
y^2) B'\Biggr\} . \qquad \\
& & \hspace{-.3cm} = 2 (D \!-\!1) \Biggl\{(D \!-\!1) I\Bigl[(2 \!-\! y) B'
\Bigr] \!-\! (D \!-\! 2) (2 \!-\! y) B' \!+\! {\rm const} \Biggr\} . \qquad
\end{eqnarray}
Acting on the right hand side of (\ref{calAgam}) and using identities
(\ref{BID2}), (\ref{BCID}) and (\ref{ABC}) proves the relation,
\begin{eqnarray}
\lefteqn{\frac{\square}{H^2} \Biggl\{\frac14 \Bigl(\frac{D \!-\! 3}{D \!-\! 1}
\Bigr) I\Bigl[(4 y \!-\! y^2) \gamma' \!+\! (D\!-\!1) (2 \!-\!y) \gamma\Bigr] - 
\frac12 (D\!-\!2) I[B] \Biggr\} } \nonumber \\
& & \hspace{.5cm} = \frac12 (D \!-\!3) (D\!-\! 1) I\Bigl[ (2 \!-\!y) B'\Bigr] 
- \frac12 (D\!-\!3) (D\!-\!2) (2 \!-\! y) B \nonumber \\
& & \hspace{7cm} - (D\!-\!2) (2 \!-\! y) B + {\rm const} \; , \qquad \\
& & \hspace{.5cm} = \frac12 (D \!-\!3) (D \!-\!1) I[B] -\frac12 (D\!-\!1) 
(2 \!-\! y) B + {\rm const} \; , \qquad \\
& & \hspace{.5cm} = -(D \!-\! 1) A + {\rm const} \; .
\end{eqnarray}

The point of enduring all this analysis is that we can now recognize the
``Integral Term'' of the invariant propagator (\ref{invprop}) as a
collection of propagators plus the $\ln^2(a_x a_z)$ term of expression
(\ref{logsq}),
\begin{eqnarray}
\mathcal{I}(x;z) & \equiv & \frac1{4 (D\!-\!1) H^2} \, I\Biggl[- (4 y \!-\! 
y^2) \gamma' \!-\! (D\!-\!1) (2 \!-\! y) \gamma \!+\! 2 (D \!-\!1) B\Biggr] ,
\label{invInt0} \\
& = & -\frac{\mathcal{A}}{(D \!-\! 3) H^2} - \Biggl[\frac{C \! - \! A}{(D 
\!-\! 3)^2 H^2} \Biggr] + {\rm const} \; , \qquad \\
& = & - \frac{i\Delta_{A'}(x;z)}{(D \!-\!3) H^2} - \Biggl[\frac{i\Delta_C(x;z) 
\!-\! i\Delta_A(x;z)}{(D \!-\! 3)^2 H^2} \Biggr] + \frac{k \ln^2(a_x a_z)}{2 
(D \!-\!3) H^2} \nonumber \\
& & \hspace{5.5cm} + {\rm const} \times \ln(a_x a_z) + {\rm const} 
\; . \qquad \label{invInt}
\end{eqnarray}
Note that the two unknown constants are irrelevant because they drop 
out when one differentiates with respect to $x^{\mu}$ and $z^{\nu}$ to 
get the propagator.

We can make contact between the Integral Term (\ref{invInt}) of the invariant
propagator and the Integral Term (\ref{Integral}) of the transformed 
propagator by expressing the propagators as convolution integrals,
\begin{eqnarray}
\lefteqn{-\frac{i\Delta_{A'}(x;z)}{(D \!-\!3) H^2} = 
\Bigl(\frac{D \!-\! 1}{D \!-\! 3}\Bigr) \!\! \int_{V} \!\! 
d^Dx' \!\sqrt{-g(x')} \, G(x;x') i\Delta_A(x';z) - \frac{\mathcal{S}_A(x;z)}{
(D \!-\!3) H^2} \, , \qquad } \label{A'term} \\
\lefteqn{-\Biggl[\frac{i\Delta_C(x;z) \!-\! i\Delta_A(x;z)}{(D \!-\! 3)^2 H^2} 
\Biggr] = - \!\! \int_{V} \!\! d^Dx' \sqrt{-g(x')} \, 
G(x;x') \frac{i\Delta_C(x';z)}{D \!-\!3} } \nonumber \\
& & - \frac{\mathcal{S}_C(x;z)}{2 (D \!-\!3)^2 H^2} - \! 
\int_{V} \!\! d^Dz' \sqrt{-g(z')} \, G(z;z') \frac{i\Delta_C(x;z')}{D \!-\!3} 
- \frac{\mathcal{S}_C(z;x)}{2 (D \!-\!3)^2 H^2} \, . \qquad  \label{Cterm}
\end{eqnarray}
Now break up the prefactor of (\ref{A'term}) as,
\begin{equation}
\Bigl(\frac{D \!-\! 1}{D \!-\! 3}\Bigr) = 1 + \frac1{D \!-\! 3} + 
\frac1{D \!-\!3} \; ,
\end{equation}
and combine the convolutions multiplying the last two factors with the 
convolutions of (\ref{Cterm}) to produce the combination $i\Delta_C -
i\Delta_A$ that can be recognized as another convolution,
\begin{eqnarray}
\lefteqn{-\frac{i\Delta_{A'}(x;z)}{(D \!-\!3) H^2} -\Biggl[\frac{i\Delta_C(x;z)
 \!-\! i\Delta_A(x;z)}{(D \!-\! 3)^2 H^2} \Biggr] } \nonumber \\
& & = -i \!\! \int_{V} \!\! d^Dx' \sqrt{-g(x')} \, i\Delta_A(x;x') 
i\Delta_A(x';z) \nonumber \\
& & \hspace{1cm} + i \!\! \int_{V} \!\! d^Dx' \sqrt{-g(x')} \, i\Delta_A(x;x')
\Biggl[ \frac{ i\Delta_C(z;x') \!-\! i\Delta_A(z;x')}{D -3} \Biggr]
\nonumber \\
& & \hspace{1cm} + i \!\! \int_{V} \!\! d^Dz' \sqrt{-g(z')} \, i\Delta_A(z;z')
\Biggl[ \frac{ i\Delta_C(x;z') \!-\! i\Delta_A(x;z')}{D -3} \Biggr]
\nonumber \\
& & \hspace{4cm} - \frac{\mathcal{S}_A(x;z)}{(D \!-\!3) H^2}
- \frac{\mathcal{S}_C(x;z)}{2 (D \!-\!3)^2 H^2} - \frac{\mathcal{S}_C(z;x)}{
2 (D \!-\!3)^2 H^2} \; , \qquad \\
& & = -i \!\! \int_{V} \!\! d^Dx' \sqrt{-g(x')} \, i\Delta_A(x;x') 
i\Delta_A(x';z) \nonumber \\
& & \hspace{.5cm} + 4 H^2 \!\! \int_{V} \!\! d^Dx' \sqrt{-g(x')} \, 
i\Delta_A(x;x') \!\! \int_{V} \!\! d^Dz' \sqrt{-g(z')} \, i\Delta_A(z;z') \,
i\Delta_C(x';z') \nonumber \\
& & \hspace{.5cm} - \!\! \int_{V} \!\! d^Dx' \sqrt{-g(x')} \, G(x;x') 
\frac{\mathcal{S}_C(z;x')}{D \!-\! 3} 
- \!\! \int_{V} \!\! d^Dz' \sqrt{-g(z')} \, G(z;z') 
\frac{\mathcal{S}_C(x;z')}{D \!-\! 3} \nonumber \\
& & \hspace{4cm} - \frac{\mathcal{S}_A(x;z)}{(D \!-\!3) H^2} - \frac{
\mathcal{S}_C(x;z)}{ 2 (D \!-\!3)^2 H^2} - \frac{\mathcal{S}_C(z;x)}{2
(D \!-\! 3)^2 H^2} \; . \qquad
\end{eqnarray}
We obtain the desired relation by adding the Integral Term (\ref{Integral}) 
of the transformed propagator to the $\ln^2(a_x a_z)$ contribution 
(\ref{logsq}) from the Other Term (and some pieces which drop out when 
differentiated by $x^{\mu}$ and $z^{\nu}$),
\begin{eqnarray}
\lefteqn{\overline{\mathcal{I}}(x;z) + \frac{k \ln^2(a_x a_z)}{2 (D\!-\!3) H^2} 
+ {\rm const} \times \ln(a_x a_z) + {\rm const} } \nonumber \\
& & \hspace{1.5cm}=\mathcal{I}(x;z) + \frac{\mathcal{S}_A(x;z)}{(D \!-\!3) H^2}
+ \Biggl[\frac{\mathcal{S}_C(x;z) \!+\! \mathcal{S}_C(z;x)}{2 (D\!-\!3)^2 H^2}
\Biggr] \nonumber \\
& & \hspace{2.5cm} + \!\! \int_{V} \!\! d^Dx' \sqrt{-g(x')} \, G(x;x') 
\Biggl[\mathcal{S}_B(z;x')  \!+\! \frac{\mathcal{S}_C(z;x')}{D \!-\! 3}
\Biggr] \nonumber \\
& & \hspace{3.5cm} + \!\! \int_{V} \!\! d^Dz' \sqrt{-g(z')} \, G(z;z') 
\Biggl[\mathcal{S}_B(x;z')  \!+\! \frac{\mathcal{S}_C(x;z')}{D \!-\! 3}
\Biggr] . \qquad \label{Ibar}
\end{eqnarray}

\section{Determining the Homogeneous Part}

The purpose of this section is to show that we can make the transformed
propagator agree with the invariant one by correctly choosing the homogeneous 
part of the full gauge parameter $\theta[A](x)$. We begin by summarizing the
relevant results of the previous two sections concerning the gauge parameter
$\overline{\theta}[A](x)$, given in (\ref{thetabar}), which enforces Lorentz 
gauge but not de Sitter invariance. The resulting propagator agrees with the
invariant one (\ref{dSprop}-\ref{gamma}) up to three surface terms which we
denote as ``$A$-type,'' ``$B$-type'' and ``$C$-type'' according to the mode
functions which they involve. We then exhibit a homogeneous gauge parameter 
$\Delta \theta[A](x)$, depending upon $A_0$, which can be used to absorb the 
$B$-type and $C$-type surface terms. The section closes by deriving a
homogeneous gauge parameter $\delta \theta[A](x)$, depending upon $A_i$, 
which absorbs the $A$-type surface terms and results in complete agreement
with the invariant propagator.

\subsection{Summary of Previous Results}

Our goal is to construct a functional change of variables that is also a 
gauge transformation,
\begin{equation}
A_{\mu}'(x) = A_{\mu}(x) - \partial_{\mu} \theta[A](x) \; .
\end{equation}
We want the field-dependent gauge parameter $\theta[A](x)$ to do two things:
\begin{enumerate}
\item{Make the field $A_{\mu}'(x)$ obey Lorentz gauge; and}
\item{Make the propagator associated with $A_{\mu}'(x)$ agree with the
unique, de Sitter invariant solution of the Lorentz gauge propagator
equation \cite{TW7}.}
\end{enumerate}
The first condition implies a second order differential equation for 
$\theta[A](x)$,
\begin{equation}
\sqrt{-g} \, \square \theta = \partial_{\mu} \Bigl[\sqrt{-g} g^{\mu\nu} 
A_{\nu}\Bigr] \; . \label{condition}
\end{equation}
Of course this only defines $\theta[A](x)$ up to a term which is 
annihilated by the scalar d'Alembertian. Because propagators obey Feynman
boundary conditions we took the inhomogeneous solution to be the convolution 
of $-i$ times the scalar propagator ($G(x;z) \equiv -i \times i\Delta_A(x;z)$)
with the right hand side of (\ref{condition}),
\begin{equation}
\overline{\theta}[A](x) \equiv \int_{V} \!\! d^Dx' \, G(x;x') 
\frac{\partial}{\partial x^{\prime \rho}} \Bigl[ \sqrt{-g(x')} \, 
g^{\rho\sigma}(x') A_{\sigma}(x')\Bigr] \; . \label{thbar}
\end{equation}
The result of just performing this transformation defines a field,
\begin{equation}
\overline{A}_{\mu}(x) \equiv A_{\mu}(x) - \partial_{\mu} 
\overline{\theta}[A](x) \; , \label{partial}
\end{equation}
which obeys the Lorentz gauge condition but whose propagator does not quite 
agree with the invariant one. 

We decomposed the propagator of $\overline{A}_{\mu}(x)$ into the double 
gradient of an ``Integral Term'' (\ref{Integral}) and an ``Other Term'' 
(\ref{Other}),
\begin{equation}
\Bigl\langle \Omega \Bigl\vert T^*\Bigl[\overline{A}_{\mu}(x) \overline{A
}_{\nu}(z) \Bigr] \Bigr\vert \Omega \Bigr\rangle = \frac{\partial}{\partial 
x^{\mu}} \frac{\partial}{\partial z^{\nu}} \, \overline{\mathcal{I}}(x;z) +
\Bigl[\mbox{}_{\mu} \overline{\mathcal{O}}_{\nu}\Bigr](x;z) \; .
\end{equation}
The invariant propagator can be broken up in similar fashion (\ref{invprop}),
\begin{equation}
i\Bigl[\mbox{}_{\mu} \Delta^{\rm dS}_{\nu}\Bigr](x;z)= \frac{\partial}{\partial
x^{\mu}} \frac{\partial}{\partial z^{\nu}} \, \mathcal{I}(x;z) - \frac1{2 H^2}
B\Bigl(y(x;z)\Bigr) \frac{\partial^2 y(x;z)}{\partial x^{\mu} \partial z^{\nu}}
\; .
\end{equation}
It is desirable to shift the double gradient of a spatially constant term,
\begin{equation}
\Delta \overline{\mathcal{I}}(x;z) \equiv \frac{k \ln^2(a_x a_z)}{2 (D \!-\!3)}
+ {\rm const} \times \ln(a_x a_z) + {\rm const} \; ,
\end{equation}
from $[\mbox{}_{\mu} \overline{\mathcal{O}}_{\nu}](x;z)$ to $\overline{
\mathcal{I}}(x;z)$. When this is done, the difference between what we want the
full transformation to produce and what the $\overline{\theta}[A](x)$ 
transformation actually gives is,
\begin{eqnarray}
\lefteqn{\Bigl[\mbox{}_{\mu} \mathcal{O}_{\nu}\Bigr] -
\Bigl[\mbox{}_{\mu} \overline{\mathcal{O}}_{\nu}\Bigr] +
\frac{\partial^2 \Delta \overline{\mathcal{I}}}{ \partial x^{\mu} 
\partial z^{\nu}} = - \frac{a_z \delta^0_{\nu}}{D \!-\!3} \frac{\partial
\mathcal{S}_C(x;z)}{\partial H x^{\mu}} - \frac{a_x \delta^0_{\mu}}{
D \!-\!3} \frac{\partial \mathcal{S}_C(z;x)}{\partial H z^{\nu}} \; , \qquad} 
\label{DelO} \\
\lefteqn{\mathcal{I}(x;z) - \overline{\mathcal{I}}(x;z) - 
\Delta \overline{\mathcal{I}}(x;z) 
= - \frac{\mathcal{S}_A(x;z)}{(D \!-\!3) H^2}
- \Biggl[\frac{\mathcal{S}_C(x;z) \!+\! \mathcal{S}_C(z;x)}{2 (D\!-\!3)^2 H^2}
\Biggr] } \nonumber \\
& & \hspace{2.5cm} - \!\! \int_{V} \!\! d^Dx' \sqrt{-g(x')} \, G(x;x') 
\Biggl[\mathcal{S}_B(z;x')  \!+\! \frac{\mathcal{S}_C(z;x')}{D \!-\! 3}
\Biggr] \nonumber \\
& & \hspace{3.5cm} - \!\! \int_{V} \!\! d^Dz' \sqrt{-g(z')} \, G(z;z') 
\Biggl[\mathcal{S}_B(x;z')  \!+\! \frac{\mathcal{S}_C(x;z')}{D \!-\! 3}
\Biggr] . \qquad \label{DelI}
\end{eqnarray}
Each of the surface integrals, $\mathcal{S}_F(x;z)$, consists of a Dirichlet 
and a Neumann contribution,
\begin{equation}
\mathcal{S}_F(x;z) \equiv \!\! \int_{\partial V}\!\!\!\!\!\! d^{D-1}\!x'_{\rho} 
\!\sqrt{-g'} g^{\prime \rho\sigma} \Bigl[F(x';z) \partial_{\sigma}' G(x;x')
\!-\! G(x;x') \partial_{\sigma}' F(x';z) \Bigr] . \label{surfF}
\end{equation}
The functions $F(x;z)$ associated with the three integrals are,
\begin{eqnarray}
\mathcal{S}_A(x;z) & \Longrightarrow & F(x';z) = i\Delta_{A'}(x';z) \; , \\
\mathcal{S}_B(x;z) & \Longrightarrow & F(x';z) = \frac{a_{x'}}{a_z}
i\Delta_B(x';z) \; , \\
\mathcal{S}_C(x;z) & \Longrightarrow & F(x';z) = i\Delta_C(x';z) \; .
\end{eqnarray}
Note that each $\mathcal{S}_F(x;z)$ is homogeneous on the first argument
$x^{\mu}$. The integral $\mathcal{S}_A(x;z)$ is homogeneous on $z^{\mu}$ as 
well, and also symmetric under interchange of $x^{\mu}$ and $z^{\mu}$.

\subsection{Absorbing the $B$-type and $C$-type Surface Terms}

Rather than absorb all the surface terms at once it is simpler to
first cancel those of the ``Other Term,'' which must also reduce those 
that remain in the ``Integral Term'' to pure $A$-type (homogeneous on
both $x^{\mu}$ and $z^{\mu}$). We accordingly seek a homogeneous gauge 
parameter $\Delta \theta[A](x)$ which cancels (\ref{DelO}). Because
this will also change the ``Integral Term'' we write out the full 
transformed field,
\begin{equation}
\widehat{A}_{\mu}(x) = \overline{A}_{\mu}(x) -\frac{\partial}{\partial x^{\mu}} 
\Delta \theta[A](x) \; .
\end{equation}
The propagator of $\widehat{A}$ is,
\begin{eqnarray}
\lefteqn{\Bigl\langle \Omega \Bigl\vert T^*\Bigl[\widehat{A}_{\mu}(x) 
\widehat{A}_{\nu}(z) \Bigr] \Bigr\vert \Omega \Bigr\rangle = 
\Bigl\langle \Omega \Bigl\vert T^*\Bigl[ \overline{A}_{\mu}(x) 
\overline{A}_{\nu}(z) \Bigr] \Bigr\vert \Omega \Bigr\rangle } \nonumber \\
& & - \frac{\partial}{\partial x^{\mu}} \Bigl\langle \Omega \Bigl\vert 
T^*\Bigl[\Delta \theta(x) A_{\nu}(z) \Bigr] \Bigr\vert \Omega \Bigr\rangle 
- \frac{\partial}{\partial z^{\nu}} \Bigl\langle \Omega \Bigl\vert T^*\Bigl[
A_{\mu}(x) \Delta \theta(z) \Bigr] \Bigr\vert \Omega \Bigr\rangle \nonumber \\
& & \hspace{1.3cm} + \frac{\partial}{\partial x^{\mu}} \frac{\partial}{\partial
z^{\nu}} \Bigl\langle \Omega \Bigl\vert T^*\Bigl[\Delta \theta(x) 
\overline{\theta}(z) \!+\! \overline{\theta}(x) \Delta \theta(z) \!+\!
\Delta \theta(x) \Delta \theta(z) \Bigr] \Bigr\vert \Omega \Bigr\rangle 
\; . \qquad \label{fullprop}
\end{eqnarray}
The terms on the final line of (\ref{fullprop}) must belong to the Integral
Term (\ref{DelI}), and  most of the middle line of (\ref{fullprop}) must 
similarly belong to the Other Term (\ref{DelO}). If we assume $\Delta 
\theta[A](x)$ depends only upon $A_0$ then the break is clean and we have,
\begin{eqnarray}
\lefteqn{ \Bigl\langle \Omega \Bigl\vert T^*\Bigl[\Delta \theta(x) A_0(z) 
\Bigr] \Bigr\vert \Omega \Bigr\rangle = \frac{a_z \mathcal{S}_C(x;z)}{(D \!-\!
3) H} \; , } \label{con1} \\
\lefteqn{\Bigl\langle \Omega \Bigl\vert T^*\Bigl[\Delta \theta(x) 
\overline{\theta}(z) \!+\! \overline{\theta}(x) \Delta \theta(z) \!+\!
\Delta \theta(x) \Delta \theta(z) \Bigr] \Bigr\vert \Omega \Bigr\rangle 
= \Bigl( A\!\!-\!\!{\rm type\ Terms} \Bigr) } \nonumber \\
& & - \Biggl[\frac{\mathcal{S}_C(x;z) \!+\! \mathcal{S}_C(z;x)}{2 (D\!-\!3)^2 
H^2} \Biggr] - \!\! \int_{V} \!\! d^Dx' \sqrt{-g(x')} \, G(x;x') 
\Biggl[\mathcal{S}_B(z;x')  \!+\! \frac{\mathcal{S}_C(z;x')}{D \!-\! 3}
\Biggr] \nonumber \\
& & \hspace{3.5cm} - \!\! \int_{V} \!\! d^Dz' \sqrt{-g(z')} \, G(z;z') 
\Biggl[\mathcal{S}_B(x;z')  \!+\! \frac{\mathcal{S}_C(x;z')}{D \!-\! 3}
\Biggr] . \qquad \label{con2} 
\end{eqnarray}

It is straightforward to see that relation (\ref{con1}) fixes the homogeneous
part of the gauge parameter to be,
\begin{equation}
\Delta \theta(x) = \frac{-1}{(D \!-\!3) H} \! \int_{\partial V} \!\!\! 
d^{D-1}\!x_{\rho}' \sqrt{-g'} g^{\prime \rho\sigma} \Biggr[ \frac{A_0(x')}{
a_{x'}} \, \partial_{\sigma}' G(x;x') - G(x;x') \partial_{\sigma}' 
\frac{A_0(x')}{a_{x'}} \Biggr] . \label{Deltheta}
\end{equation}
Combining (\ref{thbar}) and (\ref{Deltheta}) gives,
\begin{equation}
\Bigl\langle \Omega \Bigl\vert T^*\Bigl[\Delta \theta(x) \overline{\theta}(z) 
\Bigr] \Bigr\vert \Omega \Bigr\rangle = \frac{-1}{(D \!-\!3) H} \! \int_{V}\!\!
d^Dz' \, G(z;z') \frac{\partial}{\partial z^{\prime 0}} \Bigl[ a_{z'}^{D-1}
\mathcal{S}_C(x;z')\Bigr] \; .
\end{equation}
The surface integral $\mathcal{S}_C(x;z')$ has the form (\ref{surfF}) with
the function $F(x';z') = i\Delta_C(x';z')$. Multiplying this by the factor
of $a_{z'}^{D-1}$ and taking the derivative gives an expression which we can
simplify using relations (\ref{ABC}) and (\ref{BCID}),
\begin{eqnarray}
\lefteqn{ \frac{\partial}{\partial z^{\prime 0}} \Bigl[ a_{z'}^{D-1}
i\Delta_C(x';z')\Bigr] = H a_{z'}^D \Biggl\{ (D \!-\!1) C - (2 \!-\!y) C'
+ 2 \frac{a_{x'}}{a_{z'}} \, C'\Biggr\} , } \\
& & = H a_{z'}^D \Biggl\{2 C + (D \!-\! 3) C -\frac12 (D \!-\!3)
(2 \!-\!y) B \nonumber \\
& & \hspace{5cm} + (D \!-\!3) \frac{a_{x'}}{a_{z'}} \, B - (2 \!-\! y) A' + 2
\frac{a_{x'}}{a_{z'}} \, A' \Biggr\} , \qquad \\
& & = H a_{z'}^D \Biggl\{2 C + (D \!-\! 3) \frac{a_{x'}}{a_{z'}} \, B\Biggr\}
+ a_{z'}^{D-1} \frac{\partial}{\partial z^{\prime 0}} \Bigl[ i\Delta_A(x';z')
\Bigr] \; .
\end{eqnarray}
The final term involving $i\Delta_A(x';z')$ gives rise to an $A$-type surface
term whose form we will work out in the next subsection. We can therefore 
write,
\begin{eqnarray}
\lefteqn{\Bigl\langle \Omega \Bigl\vert T^*\Bigl[\Delta \theta(x) 
\overline{\theta}(z) \Bigr] \Bigr\vert \Omega \Bigr\rangle 
= \Bigl( A\!\!-\!\!{\rm type\ Terms} \Bigr) } \nonumber \\
& & \hspace{2cm}  -\!\! \int_{V} \!\! d^Dz' \sqrt{-g(z')} \, G(z;z') 
\Biggl[\mathcal{S}_B(x;z')  \!+\! \frac2{D \!-\!3} \, \mathcal{S}_C(x;z')
\Biggr] . \qquad \label{mixed1}
\end{eqnarray}
Interchanging $x^{\mu}$ and $z^{\mu}$ gives,
\begin{eqnarray}
\lefteqn{\Bigl\langle \Omega \Bigl\vert T^*\Bigl[\overline{\theta}(x) 
\Delta \theta(z) \Bigr] \Bigr\vert \Omega \Bigr\rangle 
= \Bigl( A\!\!-\!\!{\rm type\ Terms} \Bigr) } \nonumber \\
& & \hspace{2cm} -\!\! \int_{V} \!\! d^Dx' \sqrt{-g(x')} \, G(x;x') 
\Biggl[\mathcal{S}_B(z;x')  \!+\! \frac2{D \!-\!3} \, \mathcal{S}_C(z;x')
\Biggr] . \qquad \label{mixed2}
\end{eqnarray}

The term with two $\Delta \theta$'s yields a surface integral of
surface integrals that we can write as a volume integral of surface
integrals using Green's 2nd identity,
\begin{eqnarray}
\lefteqn{\Bigl\langle \Omega \Bigl\vert T^*\Bigl[\Delta \theta(x) 
\Delta \theta(z) \Bigr] \Bigr\vert \Omega \Bigr\rangle = - \frac1{(D \!-\!3)^2 
H^2} \!\! \int_{\partial V} \!\!\! d^{D-1}\!x_{\rho}' \sqrt{-g(x')} \,
g^{\rho\sigma}(x') } \nonumber \\
& & \hspace{3cm} \times \Biggl[ \mathcal{S}_C(z;x')
\frac{\partial}{\partial x^{\prime \sigma}} G(x;x') - G(x;x') \frac{\partial}{
\partial x^{\prime \sigma}} \mathcal{S}_C(z;x') \Biggr] \; , \qquad \\
& & \hspace{-.7cm} = \! \frac{-1}{(D \!-\!3)^2 H^2} \!\!\int_{V} \!\!\! d^D\!x' 
\!\sqrt{\!-g(x')} \Bigl[ \mathcal{S}_C(z;x') \square_{x'} G(x;x') \!-\! G(x;x')
\square_{x'} \mathcal{S}_C(z;x')\Bigr] . \qquad
\end{eqnarray}
Of course we can use the identity $\sqrt{-g(x')} \, \square_{x'} G(x;x') = i
\delta^D(x - x')$, and the quantity $\sqrt{-g(x')} \, \square_{x'} 
\mathcal{S}_C(z;x')$ involves,
\begin{equation}
\sqrt{-g(x')} \, \square_{x'} i\Delta_C(z';x') = i\delta^D(x' \!-\! z')
+ 2(D\!-\!3) H^2 \sqrt{-g(x')} \, i\Delta_C(z';x') \; . \label{BoxC}
\end{equation}
The delta function in (\ref{BoxC}) gives another $A$-type Term whose form 
we work out in the next subsection. Hence we have,
\begin{eqnarray}
\lefteqn{\Bigl\langle \Omega \Bigl\vert T^*\Bigl[\Delta \theta(x) 
\Delta \theta(z) \Bigr] \Bigr\vert \Omega \Bigr\rangle = 
\Bigl( A\!\!-\!\!{\rm type\ Terms} \Bigr) } \nonumber \\
& & \hspace{2cm} - \frac{\mathcal{S}_C(z;x)}{(D \!-\! 3)^2 H^2} + 
\frac2{D \!-\!3} \! \int_{V} \!\! d^Dx' \sqrt{-g(x')} \, G(x;x')
\mathcal{S}_C(z;x') \; . \qquad
\end{eqnarray}
The result is symmetric in $x^{\mu}$ and $z^{\mu}$ so we can express it as,
\begin{eqnarray}
\lefteqn{\Bigl\langle \Omega \Bigl\vert T^*\Bigl[\Delta \theta(x) 
\Delta \theta(z) \Bigr] \Bigr\vert \Omega \Bigr\rangle = 
\Bigl( A\!\!-\!\!{\rm type\ Terms} \Bigr) - 
\frac{\mathcal{S}_C(x;z) \!+\! \mathcal{S}_C(z;x)}{2 (D \!-\!3)^2 H^2} }
\nonumber \\
& & \hspace{-.7cm} + \int_{V} \!\! d^Dx' \! \sqrt{-g(x')} \, G(x;x')
\frac{\mathcal{S}_C(z;x')}{D \!-\! 3} + \!\! \int_{V} \!\! d^Dz' \! 
\sqrt{-g(z')} \, G(z;z') \frac{\mathcal{S}_C(x;z')}{D \!-\! 3} \; . 
\qquad \label{pure}
\end{eqnarray}

Combining expressions (\ref{mixed1}-\ref{mixed2}) with (\ref{pure})
gives the desired form (\ref{con2}),
\begin{eqnarray}
\lefteqn{\Bigl\langle \Omega \Bigl\vert T^*\Bigl[\Delta \theta(x) 
\overline{\theta}(z) \!+\! \overline{\theta}(x) \Delta \theta(z) \!+\!
\Delta \theta(x) \Delta \theta(z) \Bigr] \Bigr\vert \Omega \Bigr\rangle 
= \Bigl( A\!\!-\!\!{\rm type\ Terms} \Bigr) } \nonumber \\
& & - \Biggl[\frac{\mathcal{S}_C(x;z) \!+\! \mathcal{S}_C(z;x)}{2 (D\!-\!3)^2 
H^2} \Biggr] - \!\! \int_{V} \!\! d^Dx' \sqrt{-g(x')} \, G(x;x') 
\Biggl[\mathcal{S}_B(z;x')  \!+\! \frac{\mathcal{S}_C(z;x')}{D \!-\! 3}
\Biggr] \nonumber \\
& & \hspace{3.5cm} - \!\! \int_{V} \!\! d^Dz' \sqrt{-g(z')} \, G(z;z') 
\Biggl[\mathcal{S}_B(x;z')  \!+\! \frac{\mathcal{S}_C(x;z')}{D \!-\! 3}
\Biggr] . \qquad
\end{eqnarray}

\subsection{Absorbing the $A$-type Surface Terms}

We should begin this section by clarifying precisely what the $A$-type 
surface terms are. They reside entirely in the ``Integral Term,'' and
they consist of $\mathcal{S}_A/(D-3)H^2$ plus the $A$-type surface terms 
induced by the gauge parameter $\Delta \theta[A]$. We first reduce
$\mathcal{S}_A$ to a pair of temporal surface terms, then derive
similar expressions for the $A$-type surface terms from $\Delta \theta[A]$.
This will motivate our construction of the final gauge parameter $\delta 
\theta[A]$ which absorbs the $A$-type surface terms and gives full
agreement with the invariant propagator.

Recall that the surface integral $\mathcal{S}_A(x;z)$ is,
\begin{equation}
\mathcal{S}_A(x;z) \equiv \!\! \int_{\partial V}\!\!\!\!\!\! d^{D-1}\!x'_{\rho} 
\!\sqrt{-g'} g^{\prime \rho\sigma} \Bigl[i\Delta_{A'}(x';z) \partial_{\sigma}' 
G(x;x') \!-\! G(x;x') \partial_{\sigma}' i\Delta_{A'}(x';z) \Bigr] ,
\end{equation}
where $i\Delta_{A'}(x;z)$ is the derivative with respect to $\nu$ (evaluated
at $\nu = (D\!-\!1)/2$) of the Fourier mode sum,
\begin{eqnarray}
\lefteqn{i\Delta_{\nu}(x;z) = \int \!\! \frac{d^{D-1}k}{(2\pi)^{D-1}} \, 
e^{i \vec{k} \cdot (\vec{x} - \vec{z})} \Biggl\{ \theta(x^0 \!-\! z^0) 
u_{\nu}(x^0,k) u^*_{\nu}(z^0,k) } \nonumber \\
& & \hspace{6.4cm} + \theta(z^0 \!-\! x^0) u^*_{\nu}(x^0,k) u_{\nu}(z^0,k)
\Biggr\} . \qquad
\end{eqnarray}
Because $G(x;z)$ is $-i$ times the same mode sum (again evaluated at $\nu = 
(D\!-\!1)/2$) we see that the surface terms at spatial infinity make no
contribution. One can therefore express $\mathcal{S}_A(x;z)$ as a Fourier 
mode sum of temporal surface terms,
\begin{eqnarray}
\lefteqn{\mathcal{S}_A(x;z) = i \! \int \!\! \frac{d^{D-1}k}{(2\pi)^{D-1}} \, 
e^{i \vec{k} \cdot (\vec{x} - \vec{z})} } \nonumber \\
& & \hspace{-.3cm} \times \Biggl\{ u_A^*(x^0,k) u_A^*(z^0,k) \times
\mathcal{F}(-k\eta_2) - u_A(x^0,k) u_A(z^0,k) \times \mathcal{F}^*(-k\eta_1)
\Biggr\} , \qquad
\end{eqnarray}
where $\eta_1$ and $\eta_2$ are the initial and final times, respectively,
and the function $\mathcal{F}(-k\eta)$ is,
\begin{eqnarray}
\lefteqn{\mathcal{F}(-k\eta) \equiv a^{D-2} \Biggl\{ \frac{\partial 
u_{\nu}(\eta,k)}{\partial \nu} \frac{\partial u_{\nu}(\eta,k)}{\partial \eta}
- u_{\nu}(\eta,k) \frac{\partial^2 u_{\nu}(\eta,k)}{\partial \nu \partial \eta}
\Biggr\}_{\nu = \frac{D-1}2} \!\!\!\!\! , } \\
& & = \frac{\pi}{4 H a} \Biggl\{ \frac{\partial H^{(1)}_{\nu}(-k\eta)}{\partial 
\nu} \frac{\partial H^{(1)}_{\nu}(-k\eta)}{\partial \eta} - H^{(1)}_{\nu}(-k
\eta) \frac{\partial^2 H^{(1)}_{\nu}(-k\eta)}{\partial \nu \partial \eta}
\Biggr\}_{\nu = \frac{D-1}2} \!\!\!\!\! . \qquad
\end{eqnarray}
This function $\mathcal{F}(z)$ has the interesting property that it can be
related to the product of two Hankel functions, without any derivatives
with respect to the index or the argument \cite{TW9}. To see the relation
we define,
\begin{eqnarray}
\mathcal{E}_{\nu}(z) & \equiv & z \Bigl[ H^{(1)}_{\nu}(z)\Bigr]^2 \; , \\
\mathcal{G}_{\nu}(z) & \equiv & z \Biggl[ \partial_{\nu} H^{(1)}_{\nu}(z)
\partial_z H^{(1)}_{\nu}(z) - H^{(1)}_{\nu}(z) \partial_{\nu} \partial_z
H^{(1)}_{\nu}(z) \Biggr] .
\end{eqnarray}
Of course we have,
\begin{equation}
\mathcal{F}(-k\eta) = -\frac{\pi}{4} \times \mathcal{G}_{\nu}(-k\eta) \; ,
\end{equation}
and the relation to $\mathcal{E}_{\nu}$ is \cite{TW9},
\begin{equation}
\partial_z \mathcal{G}_{\nu}(z) = -\frac{2\nu}{z^2} \, \mathcal{E}_{\nu}(z)\; .
\end{equation}
The integration constant can be fixed using the asymptotic expansion for 
large $z$ to give,
\begin{equation}
\mathcal{G}_{\nu}(z) = 2 \nu \!\! \int_{z}^{\infty} \!\! dz' \,
\frac{\mathcal{E}_{\nu}(z')}{z^{\prime 2}} \; .
\end{equation}

The key identity for $\Delta \theta[A]$ to produce $A$-type surface terms is,
\begin{eqnarray}
\lefteqn{\frac{\partial}{\partial x^{\prime 0}} \frac{\partial}{\partial 
z^{\prime 0}} i\Delta_C(x';z') = \frac{i}{a_{x'}^{D-2}} \, 
\delta^D(x' \!-\! z') } \nonumber \\
& & \hspace{1cm} + \frac{\partial^2 y(x';z')}{\partial x^{\prime 0} 
\partial z^{\prime 0}} \, C'\Bigl(y(x';z')\Bigr) + 
\frac{\partial y(x';z')}{\partial x^{\prime 0}}
\frac{\partial y(x';z')}{\partial z^{\prime 0}} \, 
C''\Bigl(y(x';z')\Bigr) \; . \qquad \label{keyID}
\end{eqnarray}
The $A$-type surface terms come exclusively from the delta function term;
the other contributions produce $B$-type and $C$-type surface terms we have
already included. Note that because one gets a $D$-dimensional delta 
function, whereas the initial and final surface integrals are only 
$(D-1)$-dimensional, it is necessary to regulate $\Delta \theta[A]$ to make 
the $A$-type surface term well-defined. An obvious regularization is to 
integrate the initial and final time surfaces over a small range of duration 
$\Delta \eta = 2 \epsilon$,
\begin{eqnarray}
\lefteqn{\Delta \theta_{\epsilon}(x) \equiv \frac1{2 \epsilon (D \!-\!3) H}
\Biggl[ \int_{\eta_2 -\epsilon}^{\eta_2 + \epsilon} \!\!\!\!\!\!\!\!\!\!
dx^{\prime 0} - \int_{\eta_1 -\epsilon}^{\eta_1 + \epsilon} 
\!\!\!\!\!\!\!\!\!\!  dx^{\prime 0}\Biggr] \int \!\! d^{D-1}\!\vec{x}' }
\nonumber \\
& & \hspace{1cm} \times a_{x'}^{D-2} \Biggl\{ \frac1{a_{x'}} \, A_0(x')
\frac{\partial}{\partial x^{\prime 0}} \, G(x;x') - G(x;x')
\frac{\partial}{\partial x^{\prime 0}} \Bigl[\frac1{a_{x'}} \, A_0(x')\Bigr]
\Biggr\} . \qquad
\end{eqnarray}

Let us now work out the $A$-type surface term from $\Delta \theta_{\epsilon}(x)
\times \overline{\theta}(z)$. The full expectation value is,
\begin{eqnarray}
\lefteqn{ \Bigl\langle \Omega \Bigl\vert T^*\Bigl[ \Delta \theta_{\epsilon}(x)
\overline{\theta}(z) \Bigr] \Bigr\vert \Omega \Bigr\rangle =
\Biggl[ \int_{\eta_2 -\epsilon}^{\eta_2 + \epsilon} \!\!\!\!\!\!\!\!\!\!
dx^{\prime 0} \!-\! \int_{\eta_1 -\epsilon}^{\eta_1 + \epsilon} 
\!\!\!\!\!\!\!\!\!\!  dx^{\prime 0}\Biggr]\! \int \!\! d^{D-1}\!\vec{x}'
\!\! \int_{V} \!\! d^Dz' \frac{a_{x'}^{D-2} G(z;z')}{2 \epsilon (D \!-\!3) H}
} \nonumber \\
& & \hspace{-.5cm} \times \frac{\partial}{\partial z^{\prime 0}} 
\Biggl\{ a_{z'}^{D-1} i\Delta_C(x';z') \frac{\partial}{\partial x^{\prime 0}} 
\, G(x;x') - a_{z'}^{D-1} G(x;x') \frac{\partial}{\partial x^{\prime 0}} \, 
i\Delta_C(x';z')\Biggr\} . \qquad
\end{eqnarray}
However, we already accounted for most of this in the previous subsection; 
it is only the delta function from using (\ref{keyID}) on the final surface 
term which makes the new contribution we seek,
\begin{eqnarray}
\lefteqn{\Bigl\langle \Omega \Bigl\vert T^*\Bigl[ \Delta \theta_{\epsilon}(x)
\overline{\theta}(z) \Bigr] \Bigr\vert \Omega \Bigr\rangle_{A-{\rm type}} =
\frac{-i}{2 \epsilon (D \!-\!3) H}
\Biggl[ \int_{\eta_2 -\epsilon}^{\eta_2 + \epsilon} \!\!\!\!\!\!\!\!\!\!
dx^{\prime 0} \!-\! \int_{\eta_1 -\epsilon}^{\eta_1 + \epsilon} 
\!\!\!\!\!\!\!\!\!\!  dx^{\prime 0}\Biggr] } \nonumber \\
& & \hspace{3.5cm} \times \!\! \int \!\! d^{D-1}\!\vec{x}'
\!\! \int_{V} \!\! d^Dz' \, G(z;z') a_{z'}^{D-1} G(x;x') 
\delta^D(x' \!-\!  z') \; . \qquad
\end{eqnarray}
The integration over $z^{\prime \mu}$ is not affected by our regularization 
of $\Delta \theta_{\epsilon}(x)$,
\begin{equation}
\int \!\! d^Dz' \equiv \int_{\eta_1}^{\eta_2} \!\!\!\! dz^{\prime 0} \!
\int \!\! d^{D-1}\vec{z}' \; .
\end{equation}
It is therefore only half the $x^{\prime 0}$ range over which the delta
function can be saturated. Taking the unregulated limit gives,
\begin{eqnarray}
\lefteqn{\lim_{\epsilon \rightarrow 0} \Bigl\langle \Omega \Bigl\vert 
T^*\Bigl[ \Delta \theta_{\epsilon}(x) \overline{\theta}(z) \Bigr] \Bigr\vert 
\Omega \Bigr\rangle_{A-{\rm type}} } \nonumber \\
& & \hspace{.5cm} = \lim_{\epsilon \rightarrow 0} 
\frac{-i}{2 \epsilon (D \!-\!3) H} \Biggl[ \int_{\eta_2 -\epsilon}^{\eta_2} 
\!\!\!\!\!\!\!\!\!\! dx^{\prime 0} \!-\! \int_{\eta_1}^{\eta_1 + \epsilon} 
\!\!\!\!\!\!\!\!\!\! dx^{\prime 0}\Biggr] \! \int \!\! d^{D-1}\!\vec{x}' \,
G(z;x') a_{x'}^{D-1} G(x;x') \; , \qquad \\
& & \hspace{.5cm} = \frac{i}{2 (D\!-\!3) H} \int \!\! d^{D-1}\!\vec{x}' \,
\Biggl[ a_{x'}^{D-1} i\Delta_A(x;x') i\Delta_A(z;x') \Biggr]_{x^{\prime 0}
= \eta_1}^{x^{\prime 0} = \eta_2} . \qquad \label{left}
\end{eqnarray}

One obviously gets the same result (\ref{left}) from $\overline{\theta}(x)
\times \Delta \theta_{\epsilon}(z)$ so the total for these ``mixed'' terms
is,
\begin{eqnarray}
\lefteqn{\lim_{\epsilon \rightarrow 0} \Bigl\langle \Omega \Bigl\vert 
T^*\Bigl[ \Delta \theta_{\epsilon}(x) \overline{\theta}(z) \!+\! 
\overline{\theta}(x) \Delta \theta_{\epsilon}(z) \Bigr] \Bigr\vert 
\Omega \Bigr\rangle_{A-{\rm type}} } \nonumber \\
& & \hspace{-.5cm} = \frac{i}{(D\!-\!3) H} \int \!\! d^{D-1}\!\vec{x}' \,
\Biggl[ a_{x'}^{D-1} i\Delta_A(x;x') i\Delta_A(z;x') \Biggr]_{x^{\prime 0}
= \eta_1}^{x^{\prime 0} = \eta_2} , \qquad \\
& & \hspace{-.5cm} = \frac{i}{(D\!-\!3) H} \int \!\! \frac{d^{D-1}k}{
(2\pi)^{D-1}} \, e^{i \vec{k} \cdot (\vec{x} - \vec{z})} \Biggl\{ a_2^{D-1}
u_A^*(x^0,k) u_A^*(z^0,k) \Bigl[u_A(\eta_2,k)\Bigr]^2 \nonumber \\
& & \hspace{4.5cm} - a_1^{D-1} u_A(x^0,k) u_A(z^0,k) \Bigl[u_A^*(\eta_1,k)
\Bigr]^2 \Biggr\} , \\
& & \hspace{-.5cm} = \frac{i}{(D\!-\!3) H^2} \int \!\! \frac{d^{D-1}k}{
(2\pi)^{D-1}} \, e^{i \vec{k} \cdot (\vec{x} - \vec{z})} \Biggl\{ 
u_A^*(x^0,k) u_A^*(z^0,k) \times \frac{\pi}4 \Bigl[ H^{(1)}_{\nu}(-k\eta_2)
\Bigr]^2 \nonumber \\
& & \hspace{4.5cm} -u_A(x^0,k) u_A(z^0,k) \times \frac{\pi}4 \Bigl[ 
H^{(1)}_{\nu}(-k\eta_2) \Bigr]^{*2} \Biggr\} . \qquad \label{mixed}
\end{eqnarray}
Expression (\ref{mixed}) combines nicely with the $A$-type surface term
from $\overline{\theta}$,
\begin{eqnarray}
\lefteqn{ \frac{\mathcal{S}_A(x;z)}{(D \!-\!3) H^2}
= \frac{i}{(D\!-\!3) H^2} \int \!\! \frac{d^{D-1}k}{(2\pi)^{D-1}} \, 
e^{i \vec{k} \cdot (\vec{x} - \vec{z})} } \nonumber \\
& & \hspace{-.5cm} \times \Biggl\{ u_A^*(x^0,k) u_A^*(z^0,k) \times
\mathcal{F}(-k\eta_2) - u_A(x^0,k) u_A(z^0,k) \times \mathcal{F}^*(-k\eta_1)
\Biggr\} . \qquad
\end{eqnarray}
By partial integration we can express $\mathcal{F}(z)$ as,
\begin{eqnarray}
\mathcal{F}(z) & = & -\frac{\pi}4 \times 2\nu \int_z^{\infty} \!\!\!\! dz'
\frac1{z'} \Bigl[ H^{(1)}_{\nu}(z')\Bigr]^2 \; , \\
& = & -\frac{\pi}4 \times \Bigl[ H^{(1)}_{\nu}(z)\Bigr]^2 -\frac{\pi}4 
\times \int_z^{\infty} \!\!\!\! dz' \frac1{z^{\prime 2\nu}} 
\frac{\partial}{\partial z'} \Bigl[ z^{\prime \nu} H^{(1)}_{\nu}(z')\Bigr]^2 
\; . \qquad \label{reexp}
\end{eqnarray}
So the first term of (\ref{reexp}) is cancelled by (\ref{mixed}).

The full expectation value for $\Delta \theta_{\epsilon}(x) \times \Delta 
\theta_{\epsilon}(z)$ is,
\begin{eqnarray}
\lefteqn{\Bigl\langle \Omega \Bigl\vert T^*\Bigl[ \Delta \theta_{\epsilon}(x) 
\Delta \theta_{\epsilon}(z) \Bigr] \Bigr\vert \Omega \Bigr\rangle 
= \frac{-1}{4 \epsilon^2 (D \!-\!3)^2 H^2} } \nonumber \\
& & \hspace{-.5cm} \times \Biggl[ \int_{\eta_2 -\epsilon}^{\eta_2 + \epsilon} 
\!\!\!\!\!\!\!\!\!\! dx^{\prime 0} \!-\! \int_{\eta_1-\epsilon}^{\eta_1 + 
\epsilon} \!\!\!\!\!\!\!\!\!\! dx^{\prime 0} \Biggr] \! \int \!\! 
d^{D-1}\!\vec{x}' \, a_{x'}^{D-2} \Biggl[ \int_{\eta_2 -\epsilon}^{\eta_2 +
\epsilon} \!\!\!\!\!\!\!\!\!\! dz^{\prime 0} \!-\! \int_{\eta_1 - \epsilon}^{
\eta_1 + \epsilon} \!\!\!\!\!\!\!\!\!\! dz^{\prime 0} \Biggr] \! \int 
\!\! d^{D-1}\!\vec{z}' \, a_{z'}^{D-2} \nonumber \\
& & \hspace{0cm} \times \Biggl\{ i\Delta_C(x';z') \frac{\partial G(x;x')}{
\partial x^{\prime 0}} \frac{\partial G(z;z')}{\partial z^{\prime 0}}
- G(z;z') \frac{\partial G(x;x')}{\partial x^{\prime 0}} 
\frac{\partial i\Delta_C(x';z')}{\partial z^{\prime 0}} \nonumber \\
& & \hspace{.5cm} - G(x;x') \frac{\partial G(z;z')}{\partial z^{\prime 0}}
\frac{\partial i\Delta_C(x';z')}{\partial x^{\prime 0}} 
+ G(x;x') G(z;z') \frac{\partial^2 i\Delta_C(x';z')}{\partial x^{\prime 0}
\partial z^{\prime 0}} \Biggr\} . \qquad
\end{eqnarray}
As with the mixed term (\ref{left}) we have already reduced most of this
in the previous subsection. The only new contribution derives from the 
delta function one obtains by using (\ref{keyID}) on the final term,
\begin{eqnarray}
\lefteqn{\Bigl\langle \Omega \Bigl\vert T^*\Bigl[ \Delta \theta_{\epsilon}(x) 
\Delta \theta_{\epsilon}(z) \Bigr] \Bigr\vert \Omega \Bigr\rangle_{A\!-\!{\rm
type}} \!\!\!\!\! = \frac{-i}{4 \epsilon^2 (D \!-\!3)^2 H^2} 
\Biggl[ \int_{\eta_2 -\epsilon}^{\eta_2 + \epsilon} \!\!\!\!\!\!\!\!\!\! 
dx^{\prime 0} \!-\! \int_{\eta_1-\epsilon}^{\eta_1 + \epsilon} 
\!\!\!\!\!\!\!\!\!\! dx^{\prime 0} \Biggr] \! \int \!\! d^{D-1}\!\vec{x}' 
\, a_{x'}^{D-2} } \nonumber \\ 
& & \hspace{2.5cm} \times \Biggl[ \int_{\eta_2 -\epsilon}^{\eta_2+ \epsilon} 
\!\!\!\!\!\!\!\!\!\! dz^{\prime 0} \!-\! \int_{\eta_1- \epsilon}^{\eta_1 + 
\epsilon} \!\!\!\!\!\!\!\!\!\! dz^{\prime 0} \Biggr] \! \int \!\! 
d^{D-1}\!\vec{z}' \, G(x;x') G(z;z') \delta^D(x' \!-\!z') \; , \qquad \\
& & \hspace{-.5cm} = \frac{i}{4 \epsilon^2 (D \!-\!3)^2 H^2} 
\Biggl[ \int_{\eta_2 -\epsilon}^{\eta_2 + \epsilon} \!\!\!\!\!\!\!\!\!\! 
dx^{\prime 0} \!-\! \int_{\eta_1 - \epsilon}^{\eta_1 + \epsilon} 
\!\!\!\!\!\!\!\!\!\! dx^{\prime 0} \Biggr] \! \int \!\! d^{D-1}\!\vec{x}' 
\, a_{x'}^{D-2} i\Delta_A(x;x') i\Delta_A(z;x') \; , \qquad \\
& & \hspace{-.5cm} = \frac{i}{4 \epsilon^2 (D \!-\!3)^2 H^2} \!\int \!\!
\frac{d^{D-1}\!k}{(2\pi)^{D-1}} \, e^{i \vec{k} \cdot (\vec{x} - \vec{z})}
\Biggl\{ u_A^*(x^0,k) u_A^*(z^0,k) 
\!\! \int_{\eta_2 -\epsilon}^{\eta_2 + \epsilon} \!\!\!\!\!\!\!\!\! d\eta \,
a_{\eta}^{D-2} \Bigl[u_A(\eta,k)\Bigr]^2 \nonumber \\
& & \hspace{4cm} - u_A(x^0,k) u_A(z^0,k) 
\!\! \int_{\eta_1 -\epsilon}^{\eta_1 + \epsilon} \!\!\!\!\!\!\!\!\! d\eta \,
a_{\eta}^{D-2} \Bigl[u_A^*(\eta,k)\Bigr]^2 \Biggr\} . \qquad
\end{eqnarray}
The integrations with respect to $\eta$ can be performed, but it is not 
possible to take the unregulated limit,
\begin{eqnarray}
\lefteqn{\int_{\eta_2 -\epsilon}^{\eta_2 + \epsilon} \!\!\!\!\!\!\!\!\! 
d\eta \, a_{\eta}^{D-2} \Bigl[u_A(\eta,k)\Bigr]^2 = -\frac{\pi}4 \! 
\int_{\eta_2 -\epsilon}^{\eta_2 + \epsilon} \!\!\!\!\!\!\!\!\! d\eta \,
\eta \Bigl[H^{(1)}_{\nu}(-k\eta)\Bigr]^2 \; , } \\
& & \hspace{1cm} = -\frac{\pi}{4 k^2} \! \int_{-k(\eta_2 -\epsilon)}^{-k(\eta_2 
+ \epsilon)} \!\!\!\!\!\!\!\!\! dz \, z \Bigl[H^{(1)}_{\nu}(z)\Bigr]^2 \; , \\
& & \hspace{1cm} = -\frac{\pi}{8 k^2} \Biggl\{ (z^2 \!-\! \nu^2) \Bigl[ 
H^{(1)}_{\nu}(z) \Bigr]^2 + z^2 \Bigl[ \frac{\partial}{\partial z} 
H^{(1)}_{\nu}(z)\Bigr]^2 \Biggr\} \Biggr\vert_{z = -k (\eta_2-\epsilon)}^{z 
= -k (\eta_2 + \epsilon)} , \qquad \\
& & \hspace{1cm} = -\frac{\pi}4 \times 2\epsilon \times \eta_2 \Bigl[ 
H^{(1)}_{\nu}(-k\eta_2) \Bigr]^2 + O(\epsilon^3) \; . \qquad 
\end{eqnarray}

Combining the various $A$-type surface terms gives a result of the form,
\begin{eqnarray}
\lefteqn{\Bigl( A\!-\!{\rm type\ Terms} \Bigr) = \frac{-i}{(D \!-\!3) H^2} 
\!\int \!\!  \frac{d^{D-1}\!k}{(2\pi)^{D-1}} \, e^{i \vec{k} \cdot (\vec{x} 
- \vec{z})} } \nonumber \\
& & \hspace{.3cm} \times \Biggl\{ u_A^*(x^0,k) u_A^*(z^0,k) 
\mathcal{A}(\eta_2,k,\epsilon) - u_A(x^0,k) u_A(z^0,k) 
\mathcal{A}^*(\eta_1,k,\epsilon) \Biggr\} . \qquad \label{finA}
\end{eqnarray}
The function $\mathcal{A}(\eta,k,\epsilon)$ is,
\begin{equation}
\mathcal{A}(\eta,k,\epsilon) = \frac{\pi}4 \Biggl\{ 
\frac{ \eta [H^{(1)}_{\nu}(-k\eta)]^2}{2 \epsilon (D\!-\!3)} +
\int_{-k\eta}^{\infty} \!\!\!\!\!\!\!\! dz' \, \frac1{z^{\prime 2\nu}}
\frac{\partial}{\partial z'} \Bigl[ z^{\prime \nu} H^{(1)}_{\nu}(z')\Bigr]^2
+ O(\epsilon) \Biggr\} . \label{scriptA}
\end{equation}
We seek a homogeneous gauge parameter $\delta \theta[A](x)$ which cancels 
(\ref{finA}-\ref{scriptA}), in the limit that $\epsilon$ goes to zero,
without changing the ``Other Term.'' If we construct it from $A_i$,
rather than $A_0$ then there will be no interference between $\Delta 
\theta[A]$ and $\delta \theta[A]$. Suppose further that $\delta \theta[A](x)$,
like $\Delta \theta[A](x)$ involves an integral over a dummy variable 
$x^{\prime \mu}$, and that the field $A_i(x')$ is differentiated with 
respect to $x^{\prime 0}$. What we want is that the expectation value of
$\delta \theta[A](x) \times A_i(z)$ is zero unless there is also a
derivative with respect to $z^0$. If we can construct a $\delta \theta[A]$
with this property then the only nonzero contribution to the transformed
propagator will come from $\delta \theta[A](x) \times \delta \theta[A](z)$.

It is simplest to construct the term we want by analogy with the simple
harmonic oscillator, whose Heisenberg position operator is,
\begin{equation}
q(t) = \frac1{\sqrt{2m\omega}} \Bigl[ a \, e^{-i\omega t} + a^{\dagger}
e^{i \omega t}\Bigr] \; . \label{SHOq}
\end{equation}
Note that we can isolate the raising and lower operators by taking
linear combinations of $\dot{q}$ and $i\omega q$,
\begin{equation}
\dot{q}(t) + i \omega q(t) = \frac{2i\omega}{\sqrt{2m\omega}} \,
a^{\dagger} e^{i\omega t} \qquad , \qquad \dot{q}(t) - i \omega q(t) 
= \frac{-2i\omega}{\sqrt{2m\omega}} \, a\, e^{-i\omega t} \; .
\label{isolate}
\end{equation}
We assume the usual commutation relations and ground state $\vert \Omega 
\rangle$,
\begin{equation}
[a,a^{\dagger}] = 1 \qquad , \qquad a \vert \Omega \rangle = 0 = \langle 
\Omega \vert a^{\dagger} \; . \label{SHOaO}
\end{equation}
If $t$ comes before the last time $t_2$, and after the earliest time $t_1$,
then we have,
\begin{eqnarray}
\Bigl\langle \Omega \Bigl\vert T^*\Bigl[ \Bigl(\dot{q}(t_2) \!+\! i \omega
q(t_2)\Bigr) q(t)\Bigr] \Bigr\vert \Omega \Bigr\rangle & = & 0 \qquad
\forall \; t< t_2 \; , \\
\Bigl\langle \Omega \Bigl\vert T^*\Bigl[ \Bigl(\dot{q}(t_1) \!-\! i \omega
q(t_1)\Bigr) q(t)\Bigr] \Bigr\vert \Omega \Bigr\rangle & = & 0 \qquad
\forall \; t_1< t \; .
\end{eqnarray}
The only nonzero expectation value comes from the $T^*$-ordered product of
two factors of the combinations (\ref{isolate}),
\begin{eqnarray}
\Bigl\langle \Omega \Bigl\vert T^*\Bigl[ \Bigl(\dot{q}(t_2) \!+\! i \omega
q(t_2)\Bigr) \Bigl(\dot{q}(t_2') \!+\! i \omega q(t_2')\Bigr) \Bigr] 
\Bigr\vert \Omega \Bigr\rangle & = & \frac{i}{m} \, \delta(t_2 \!-\! t_2') 
\; , \qquad \\
\Bigl\langle \Omega \Bigl\vert T^*\Bigl[ \Bigl(\dot{q}(t_1) \!-\! i \omega
q(t_1)\Bigr) \Bigl(\dot{q}(t_1') \!-\! i \omega q(t_1')\Bigr) \Bigr] 
\Bigr\vert \Omega \Bigr\rangle & = & -\frac{i}{m} \, \delta(t_1 \!-\! t_1') 
\; . \qquad
\end{eqnarray}

We construct a gauge parameter $\delta \theta[A](x)$ with the desired 
properties by analogy. The free field mode sum for $A_i(x)$ is \cite{RPW1,KW2},
\begin{equation}
A_i(x) = \int \!\! \frac{d^{D-1}k}{(2\pi)^{D-1}} \Biggl\{ a _{x} u_B(x^0,k)
e^{i \vec{k} \cdot \vec{x}} \beta_i(\vec{k}) + a_{x} u_B^*(x^0,k)
e^{-i \vec{k} \cdot \vec{x}} \beta_i^{\dagger}(\vec{k}) \Biggr\} ,
\end{equation}
where the $u_B(\eta,k)$ mode functions are given by (\ref{unu}) with
index $\nu = (D-3)/2$ and the $\beta_i(\vec{k})$ are canonically
normalized annihilation operators. One can isolate $\beta_i(\vec{k})$ by
taking the spatial Fourier transform,
\begin{equation}
\widetilde{A}_i(x^0,\vec{k}) \equiv \int \!\! d^{D-1}\vec{x} \, e^{-i
\vec{k} \cdot \vec{x}} A_i(x^0,\vec{x}) \; .
\end{equation}
Now form linear combinations analogous to (\ref{isolate}),
\begin{eqnarray}
\frac1{a} \, \widetilde{A}_i(\eta,\vec{k}) \frac{\partial}{\partial \eta}
\, u_B^*(\eta,k) - u_B^*(\eta,k) \frac{\partial}{\partial \eta} \, 
\Bigl[\frac1{a} \, \widetilde{A}_i(\eta,\vec{k}) \Bigr] & = & 
\frac{i \beta_i(\vec{k})}{a^{D-2}} \; , \qquad \\
\frac1{a} \, \widetilde{A}_i(\eta,-\vec{k}) \frac{\partial}{\partial \eta}
\, u_B(\eta,k) - u_B(\eta,k) \frac{\partial}{\partial \eta} \, 
\Bigl[\frac1{a} \, \widetilde{A}_i(\eta,-\vec{k}) \Bigr] & = & 
-\frac{i \beta_i^{\dagger}(\vec{k})}{a^{D-2}} \; . \qquad
\end{eqnarray}
It follows that the desired gauge parameter is,
\begin{eqnarray}
\lefteqn{ \delta \theta_{\epsilon}(x)} \nonumber \\
& & \hspace{-.5cm} = \frac{i}{\sqrt{2\epsilon}\,(D\!-\!3) H}
\! \int \!\! \frac{d^{D-1}k}{(2\pi)^{D-1}} \, e^{i \vec{k} \cdot \vec{x}} 
\Biggl\{ u_A^*(x^0,k) \! \int_{\eta_2 -\epsilon}^{\eta_2 + \epsilon} 
\!\!\!\!\!\!\!\!\!\!  dx^{\prime 0} a_{x'}^{D-2} \Bigl[(D \!-\!3)
\mathcal{A}(x^{\prime 0},k,\epsilon) \Bigr]^{\frac12} \nonumber \\
& & \hspace{1.4cm} \times \Biggl[ \frac{k_i \widetilde{A}_i(x^{\prime 0},
\vec{k})}{k a_{x'}} \frac{\partial u_B(x^{\prime 0},k)}{\partial x^{\prime 0}} 
- u_B(x^{\prime 0},k) \frac{\partial}{\partial x^{\prime 0}} \Bigl[ 
\frac{k_i \widetilde{A}_i(x^{\prime 0},\vec{k})}{k a_{x'}}\Bigr] \Biggr]
\nonumber \\
& & \hspace{0cm} - u_A(x^0,k) \! \int_{\eta_1 -\epsilon}^{\eta_1 + \epsilon} 
\!\!\!\!\!\!\!\!\!\!  dx^{\prime 0} a_{x'}^{D-2} \Bigl[(D\!-\!3)
\mathcal{A}^*(x^{\prime 0},k,\epsilon) \Bigr]^{\frac12} \nonumber \\
& & \hspace{1.4cm} \times \Biggl[ \frac{k_i \widetilde{A}_i(x^{\prime 0},
\vec{k})}{k a_{x'}}\frac{\partial u_B^*(x^{\prime 0},k)}{\partial x^{\prime 0}} 
- u_B^*(x^{\prime 0},k) \frac{\partial}{\partial x^{\prime 0}} \Bigl[ 
\frac{k_i \widetilde{A}_i(x^{\prime 0},\vec{k})}{k a_{x'}}\Bigr] \Biggr] 
\Biggr\} . \qquad
\end{eqnarray}

\section{Discussion}

There are two generic ways to freeze local symmetries:
\begin{itemize}
\item{{\it Exact Gauge Fixing}, in which the fields are made to obey some
equation; and}
\item{{\it Average Gauge Fixing}, in which a term is added to the Lagrangian.}
\end{itemize}
We have shown that certain average gauges cannot be derived from the 
canonical formalism on manifolds such as de Sitter for which there are 
linearization instabilities. Ignoring this problem in electrodynamics 
causes the vector potential to possess an unphysical and incorrect part 
which drops out of the field strength but affects interaction energies. 
This may be the origin of the on-shell singularities found in Feynman 
gauge for the one loop self-mass-squared of charged scalars on de Sitter 
\cite{KW2}.

We have also constructed the field-dependent gauge transformation that
enforces exact, Lorentz gauge on de Sitter electrodynamics. This was 
applied to the photon propagator from a non-de Sitter invariant, average 
gauge and the result agrees exactly with the de Sitter invariant solution 
previously obtained from solving the Lorentz gauge propagator equation 
\cite{TW7}. It was already known from adding the compensating gauge 
transformation to the naive de Sitter transformation that the propagator 
in the non-invariant gauge shows no physical breaking of de Sitter 
invariance \cite{RPW1}. So the fact that our transformation technique
produces an invariant result demonstrates that the technique indeed
eliminates unphysical breaking of de Sitter invariance.

In a subsequent work we will employ the same technique to transform the 
graviton propagator from a non-de Sitter invariant, average gauge 
\cite{TW3,RPW1} to the exact and de Sitter invariant, de Donder gauge.
Adding the compensating transformation shows that the breaking of de
Sitter invariance in this propagator is physical \cite{Kleppe}, so the
expectation is that the transformation technique will not remove it.
Because the transformed propagator will obey a de Sitter invariant gauge
condition, this should settle the issue about whether or not free 
gravitons have any de Sitter invariant states. Note that simply obeying 
a de Sitter invariant propagator equation does not guarantee a de Sitter
invariant solution, as the case of the massless, minimally coupled 
scalar proves \cite{AF}. Note also that physical graviton modes obey 
precisely the same equation as the massless, minimally coupled scalar
\cite{Grishchuk}.

Constructing the de Donder gauge propagator is a worthy goal in its
own right for two reasons. First, exploiting the gauge condition makes a
vast simplification in tensor algebra \cite{KW1}. Second, using a de
Sitter invariant gauge would preclude the need for noninvariant 
counterterms, even though the actual propagator is not de Sitter 
invariant \cite{MW,KW1}.

A significant technical result of this paper is the ``Convolution Identity''
(\ref{convolution}) for integrating the propagator $i\Delta_A$ of a massless, 
minimally coupled scalar up against the propagator $i\Delta_{\nu}$ of a 
massless scalar with conformal coupling,
\begin{equation}
\xi = \frac1{D (D \!-\!1)} \Bigl[ \Bigl(\frac{D \!-\!1}2\Bigr)^2 - \nu^2
\Bigr] \; .
\end{equation}
The result follows from Green's second identity,
\begin{eqnarray}
\lefteqn{-i \!\! \int_{V} \! d^Dx' \sqrt{-g(x')} \, i\Delta_A(x;x') 
i\Delta_{\nu}(x';z) = \frac{i\Delta_{\nu}(x;z) \!-\! i\Delta_A(x;z)}{[(
\frac{D-1}2)^2 \!-\! \nu^2] H^2} } \nonumber \\
& & \hspace{-.7cm} -i \!\! \int_{\partial V} \!\!\!\! d^{D-1}\!x'_{\rho} 
\sqrt{-g'} \, g^{\prime \rho\sigma} \Biggl[\frac{i\Delta_{\nu}(x';z) 
\partial_{\sigma}' i\Delta_A(x;x')\!-\!i\Delta_A(x;x') \partial_{\sigma}'
i\Delta_{\nu}(x';z)}{[(\frac{D-1}2)^2 \!-\! \nu^2] H^2} \Biggr] \! . \qquad 
\end{eqnarray}
We expect this to be of great utility in the subsequent graviton project
because field dependent gauge transformations result in precisely such
convolutions.

 \centerline{\bf Acknowledgements}

This work was partially supported by FQXi Mini Grant \#MGB-08-008,
by FOM grant 07PR2522, by Utrecht University, by European Union grant
MRTN-CT-2004-512194, by Hellenic grant INTERREG IIIA, by NSF grants
PHY-0653085 and PHY-0855021, and by the Institute for Fundamental Theory 
at the University of Florida.

\end{document}